\documentclass[conference]{IEEEtran}
\usepackage{mdframed}
\usepackage{tikz}
\usetikzlibrary{tikzmark,fit}
\usepackage{fbox}
\usepackage{float}
\usepackage{xcolor}
\usepackage{framed}
\usepackage{listings}
\let\labelindent\relax
\usepackage{enumitem}
\usepackage{tikz}
\usepackage{amsmath}
\usepackage{filecontents}
\usepackage[font=small,skip=0pt]{caption}
\usepackage{multirow}
\usepackage{csquotes}
\usepackage{placeins}
\usepackage{xspace}
\usepackage[normalem]{ulem}
\usepackage{soul}
\usepackage{url}
\usepackage{subcaption}
\usepackage[most]{tcolorbox}
\usepackage{listings}
\usepackage{breakurl}
\usepackage{hyperref}
\hypersetup{colorlinks,allcolors=black}    
\usepackage{enumitem}
\usepackage{xcolor}
\usepackage{mdframed}
\usepackage{tcolorbox}
\usepackage{tikz}
\usepackage{underscore}
\usepackage{graphicx}
\usepackage{subcaption}
\usepackage[most]{tcolorbox}
\usepackage{pifont}           
\usepackage{amssymb}         
\usepackage{bbding}          
\usepackage{xcolor} 
\definecolor{lightgreen}{rgb}{0.5, 0.9, 0.2}
\colorlet{mygray}{black!30}
\colorlet{mygreen}{green!90!blue}
\colorlet{mymauve}{red!60!blue}
\lstdefinelanguage{diff}{
  morecomment=[f][\color{blue}]{@@},     
  morecomment=[f][\color{red}]-,         
  morecomment=[f][\color{mygreen}]+,       
  morecomment=[f][\color{magenta}]{---}, 
  morecomment=[f][\color{magenta}]{+++},
}

\newif\ifusegreen
\usegreenfalse
\definecolor{mildgreen}{rgb}{0.0, 0.5, 0.0}

\DeclareCaptionType{example}[Example][List of Examples]
\newcommand{\tool}{\textsc{$\text{SliceLM}$}\xspace}
\newcommand{\system}{\textsc{$\text{DualLM}$}\xspace}
\newcommand{\difficult}{patches without hints}
\newcommand{\easy}{patches with hints}

\newcommand{\etal}{\emph{et al.}\xspace}
\newcommand{\cut}[1]{}

\usepackage{xstring}
\newcommand{\PP}[1]{
\vspace{2px}
\noindent{\bf \IfEndWith{#1}{.}{#1}{#1.}}
}

\newcommand{\squishlist}{
  \begin{list}{$\bullet$}
    { \setlength{\itemsep}{0pt}      \setlength{\parsep}{3pt}
      \setlength{\topsep}{3pt}       \setlength{\partopsep}{0pt}
      \setlength{\leftmargin}{1.0em} \setlength{\labelwidth}{1em}
      \setlength{\labelsep}{0.5em} } }
\newcommand{\squishend}{
    \end{list}  }

\begin{document}
\title{What Do They Fix? LLM-Aided Categorization of Security Patches for Critical Memory Bugs}

\author{%
\IEEEauthorblockN{%
Xingyu Li\textsuperscript{\textasteriskcentered},
Juefei Pu\textsuperscript{\textasteriskcentered},
Yifan Wu\textsuperscript{\textasteriskcentered},
Xiaochen Zou\textsuperscript{\textasteriskcentered},
Shitong Zhu\textsuperscript{\textasteriskcentered},
Qiushi Wu\textsuperscript{\textsection},
Zheng Zhang\textsuperscript{\textasteriskcentered},
Joshua Hsu\textsuperscript{\textasteriskcentered},\\[-0.2ex]
Yue Dong\textsuperscript{\textasteriskcentered},
Zhiyun Qian\textsuperscript{\textasteriskcentered},
Kangjie Lu\textsuperscript{\textdagger},
Trent Jaeger\textsuperscript{\textasteriskcentered},
Michael De Lucia\textsuperscript{\textdaggerdbl},
Srikanth V. Krishnamurthy\textsuperscript{\textasteriskcentered}}
\IEEEauthorblockA{\itshape\small
\makebox[\linewidth][c]{%
\textsuperscript{\textasteriskcentered}\,UC Riverside \enspace
\textsuperscript{\textdagger}\,University of Minnesota \enspace
\textsuperscript{\textsection}\,IBM \enspace
\textsuperscript{\textdaggerdbl}\,U.S.\ Army Research Laboratory%
}}
\IEEEauthorblockA{\itshape\footnotesize
\makebox[\linewidth][c]{%
\textsuperscript{\textasteriskcentered}\,\{xli399,jpu007,fshal003,xzou017,szhu014,zzhan173,jhsu094,yued,trentj,zhiyun.qian,krish\}@ucr.edu%
}\\
\makebox[\linewidth][c]{%
\textsuperscript{\textdagger}\,kjlu@umn.edu \enspace
\textsuperscript{\textsection}\,Qiushi.Wu@ibm.com \enspace
\textsuperscript{\textdaggerdbl}\,michael.j.delucia2.civ@army.mil%
}}
}

\maketitle

\begin{abstract}
 
Open-source software projects are foundational to modern software ecosystems, with the Linux kernel standing out as a critical exemplar due to its ubiquity and complexity. 
Although security patches are continuously integrated into the Linux mainline kernel, downstream maintainers often delay their adoption, creating windows of vulnerability. A key reason for this lag is the difficulty in identifying security-critical patches, particularly those addressing exploitable vulnerabilities such as out-of-bounds (OOB) accesses and use-after-free (UAF) bugs. This challenge is exacerbated by intentionally silent bug fixes, incomplete or missing CVE assignments, delays in CVE issuance, and recent changes to the CVE assignment criteria for the Linux kernel.

Prior efforts such as GraphSPD, have proposed binary classifiers to distinguish security versus non-security patches. However, these approaches do not provide fine-grained categorization of vulnerability types, which is essential for prioritizing fixes for high-impact bugs like OOB and UAF. 
Our work aims to take such coarsely labeled security patches and classify them into fine-grained categories, i.e., OOB, UAF, or non-OOB-UAF types.

While fine-grained patch classification approaches exist, they exhibit limitations in both coverage and accuracy. In this work, we identify previously unexplored opportunities to significantly improve fine-grained patch classification.
Specifically, by leveraging cues from commit titles/messages and diffs alongside appropriate code context, we develop \system, a dual-method pipeline that integrates two approaches based on a Large Language Model (LLM) and a fine-tuned small language model. \system achieves \textbf{87.4\%} accuracy and an F1-score of \textbf{0.875}, significantly outperforming prior solutions.
Notably, \system successfully identified 111 of 5,140 recent Linux kernel patches as addressing OOB or UAF vulnerabilities, with 90 true positives confirmed by manual verification (many do not have clear indications in patch descriptions). Moreover, we constructed proof-of-concepts for two identified bugs (one UAF and one OOB), including one developed to conduct a previously unknown control-flow hijack as further evidence of the correctness of the classification.

\end{abstract}

\section{Introduction}

Open-source software projects have emerged as cornerstones of modern software development ecosystems. 
However, incorporating such projects leads to significant security risks. Memory safety vulnerabilities are particularly notable due to their prevalence and severity. However, \emph{memory safety vulnerabilities vary significantly in severity}. Among them, out-of-bounds (OOB) access and use-after-free (UAF) vulnerabilities (including double-free and invalid-free) are especially perilous, as malicious actors consistently exploit them to achieve privilege escalation and compromise system integrity~\cite{exploit1, exploit2, exploit3, exploit4}. In fact, our analysis of publicly available exploits~\cite{kernelexploitdb} since 2020 showed that almost \textbf{90\%} of them exploit OOB or UAF vulnerabilities.

Taking the Linux kernel as an example, it has been shown that even when patches addressing such vulnerabilities are promptly integrated into the upstream mainline kernel, downstream kernel maintainers (e.g., Ubuntu) may delay their adoption for weeks or even months~\cite{li2024investigation,zhang2021investigation}. This extended lag creates a substantial window of vulnerability in which attackers may exploit bugs to attack unpatched downstreams. Moreover, studies have demonstrated the feasibility of automated exploit generation targeting OOB and UAF bugs~\cite{chen2020koobe,wu2018fuze,wang2018revery,xu2023autopwn}, further underscoring the urgent need for timely patch porting.

Prior work (e.g.,~\cite{zhang2021investigation}) has documented this issue extensively, which identifies a major cause of this delay: downstream maintainers often lack clear, timely, and fine-grained information about which patches are security-critical, i.e., OOB or UAF bugs~\cite{zhang2021investigation,syzscope}. The volume of daily commits makes it infeasible to manually audit every patch. Furthermore, CVE-based indicators are insufficient: (1) many security patches are fixed silently without any indications (i.e., perhaps no one understood the security impact); (2) CVE assignments often lag behind the availability of patches by weeks or even months~\cite{li2017large}; (3) CVEs are incomplete --- many exploitable security vulnerabilities 
are not assigned CVE numbers
\cite{CVEproblems,syzscope,grebe,syzbridge}.

While several tools have been proposed to detect security-related patches (e.g., VulFixMiner~\cite{zhou2021finding}, GraphSPD~\cite{wang2022graphspd}), they generally offer only binary classification (security vs. non-security), which is insufficient for practical prioritization, as not all vulnerabilities are equally impactful and urgent to be ported.
To enable downstream maintainers to triage and prioritize critical patches earlier, a more useful analysis involves a \emph{fine-grained classification} that distinguishes between vulnerability types, particularly to tease out those with high exploitability such as UAF and OOB. Although several approaches have previously been proposed, all exhibit significant limitations in both coverage and accuracy. 

\textbf{Limitations of Prior Art.}
Specifically, state-of-the-art solutions SID~\cite{wu2020precisely}, TreeVul~\cite{pan2023fine} and CoLeFunDa~\cite{zhou2023colefunda} offer fine-grained patch classifications. SID~\cite{wu2020precisely} relies on human-defined, hard-coded patterns, but the complexity and variety of patterns in the real world make them difficult to define and accurately capture. Specifically, it supports only a single hard-coded patch pattern for each UAF and OOB vulnerability type, resulting in low coverage, and fails to identify 57\% of the relevant patches.
TreeVul~\cite{pan2023fine} and CoLeFunda~\cite{zhou2023colefunda} eliminate the need for hard-coded patterns by feeding the code diffs (e.g., added and removed lines of code) to a machine-learning model. 
Unfortunately, TreeVul does not use any code context beyond the diff itself, limiting the model's ability to appropriately learn patch patterns. While ColeFunda does use code context, its use of the standard slicing technique can introduce bloated code blocks and noise.
Furthermore, both approaches suffer from a common limitation: they do not leverage commit messages, missing valuable semantic cues in natural language that can aid classification.\footnote{CoLeFunDa has three downstream tasks, what we refer for it in the paper is its CWE classification task. Also, CoLeFunDa is not open-sourced.}
Our evaluations show that (see \S\ref{sec:eval}), TreeVul achieves an accuracy of 65.6 \% and an F1-score of 0.653, in identifying patches that fix UAF and OOB bugs, indicating ample room for improvement.
These limitations motivate the need for a more effective solution that advances the state of the art in vulnerability type classification.

\textbf{Our solution.}
To tackle the challenge of handling diverse patch patterns, we propose a novel machine-learning-based solution guided by two key insights:
First, beyond code diffs in a patch, the commit titles and messages often provide valuable insights into the nature of the bug being fixed (e.g., directly mentioning a ``use-after-free'' bug being fixed). 
The advancements in LLMs make them very suitable for extracting such indicators from commit descriptions.
Different from CoLeFunDa~\cite{zhou2023colefunda} and TreeVul~\cite{pan2023fine}, our approach leverages Large Language Models (LLMs) to effectively extract and utilize these valuable indicators from commit descriptions.
While an LLM-only pipeline already outperforms recent approaches such as SID and TreeVul, its accuracy still falls short of our expectations.
Second, critical clues are often found in the code context that extends beyond the modified lines in a patch that can help identify the type of bug.
We develop a custom slicing method that concisely captures the impact of the patch to drive a fine-tuned small language model (since our slices are custom and specialized). Finally, we develop a dual-method pipeline, namely \system, that runs end-to-end automatically to classify patches into UAF, OOB, and other non-UAF-OOB patches, leveraging both types of information in an informed way.
We note that our approach differs from prior approaches relying on a single machine learning model, including CoLeFunDa~\cite{zhou2023colefunda} and TreeVul~\cite{pan2023fine}. 
In contrast, our method adopts a hybrid design that strategically integrates a large language model (LLM) with a small fine-tuned language model, effectively harnessing the strengths of both to achieve significantly higher classification performance.

We evaluated \system primarily on Linux kernel patches, achieving an 87.4\% accuracy and a 0.875 F1-score on quality-controlled CVE patches, outperforming SID and TreeVul by \textbf{23.6\%} and \textbf{21.2\%} in accuracy, and 0.329 and 0.216 in F1-score.
Most importantly,
out of 5,140 recent patches, \system identified 111 as fixing OOB or UAF bugs, with manual verification confirming \textbf{90} of these identifications as true positives.
This highlights the significant coverage of \system in teasing out critical security patches.
Notably, we construct proof-of-concepts for two such bugs (one UAF and one OOB), as further evidence of the correctness of the classification. We even successfully exploited one such bug to realize a control-flow hijack attack that was not publicly known.

\textbf{Scope and assumptions.} To show the generality of \system, we also tested our method against other types of bugs (e.g., NULL pointer dereference, memory leak, and use-before-initialization) and additional open-source projects, yielding similarly robust results.
Consistent with TreeVul and CoLeFunDa, 
we assume the patches fed to our fine-grained classifier are already security patches (e.g., CVE patches). If they are not already labeled as such, we envision a full pipeline to run existing coarse-grained classifiers that differentiate security and non-security patches, e.g., VulFixMiner~\cite{zhou2021finding} and GraphSPD~\cite{wang2022graphspd}. This is different from SID which does not require the input to be security patches.

\textbf{Contributions.} A summary of our contributions is as follows:
\squishlist 
\item We develop \system, a system that strategically leverages the strengths of an LLM and a specialized small language model to effectively classify security patches addressing critical memory corruption bugs, specifically targeting OOB and UAF vulnerabilities.
\item Our solution is driven by missed opportunities and limitations in state-of-the-art solutions, as well as recent advances in large language models, including leveraging an LLM to enhance vulnerability context extraction during slicing.
\item We show that \system achieves an 87.4\% accuracy and a 0.875 F1-score on quality-controlled CVE patches, markedly outperforming state-of-the-art methods~\cite{wu2020precisely,pan2023fine}, and identifies 90 OOB and UAF patches from 5,140 recent patches. From these, we further develop two PoCs and one exploit to achieve control flow hijacking. 
\item We will open source the code, data, and model produced as part of the research to facilitate the reproduction of the results and further research.
A draft version of the source code along with PoCs and an exploit are uploaded under ``additional materials''. 
\squishend

\section{Background and Related work}
\label{sec:background}
\subsection{Memory Corruption Vulnerabilities}

Memory corruption vulnerabilities remain among the most critical security threats in modern software systems, as they can enable attackers to achieve arbitrary code execution and escalate privileges. These vulnerabilities can arise at any stage of a memory object's lifecycle—from allocation and initialization to usage and eventual deallocation—creating multiple  opportunities for exploitation.

Among memory corruption vulnerabilities, out-of-bounds access and use-after-free are particularly critical due to their high exploit potential for arbitrary code execution and privilege escalation~\cite{wu2018fuze,chen2020koobe,wang2018revery,xu2023autopwn}. 
Our analysis of 69 publicly available exploits~\cite{kernelexploitdb} targeting Linux/Android systems since 2020 reveals that 38 were UAF vulnerabilities and 24 were OOB vulnerabilities, with only 7 being non-UAF-OOB bugs.

Security vulnerabilities are typically documented using the Common Vulnerabilities and Exposures (CVEs)~\cite{cve-intro} system. Each CVE is typically categorized using the Common Weakness Enumeration (CWE), a standardized taxonomy that classifies different types of software weaknesses and vulnerabilities. For example, CWE-787 represents the category of out-of-bounds write vulnerabilities where programs write beyond the bounds of allocated memory regions, while CWE-416 represents use-after-free vulnerabilities where programs attempt to use memory after it has been freed.

\subsection{Mining Patches}

Several studies~\cite{hoang2019patchnet,wen2019ptracer,tian2012identifying} have used machine learning to assess whether a commit in the Linux mainline is a patch or a functional change, useful for patch porting decisions for stable/LTS branches.
Once bug-fixing patches have been identified, the next challenge is to determine if such patches are security related.
Many studies~\cite{zhou2017automated,zhou2021spi,zuo2023commit,wang2019detecting,wang2021patchrnn,ganz2023pavudi,sawadogo2020learningcatchsecuritypatches,he2023bingoidentifyingsecuritypatches,zhou2021finding,nguyen2023multigranularitydetectorvulnerabilityfixes,tang2024justintimedetectionsilentsecurity}  have used machine learning to analyze patches to determine whether they are security related.
For example, VulFixMiner~\cite{zhou2021finding} employs a transformer-based model and GraphSPD~\cite{wang2022graphspd} leverages a graph neural network to perform coarse-grained patch classification (e.g., security vs. non-security). However, neither approach identifies the specific bug type, which is often critical for effective patch triaging~\cite{wu2020precisely,zhou2023colefunda}.

Some works (e.g., VFCFinder~\cite{dunlap2024vfcfinder}) focus on mapping vulnerabilities to their corresponding patch commits. However, this assumes the availability of vulnerability disclosures or CVEs, which is not always the case—especially for silent fixes. Our goal is fundamentally different: we aim to proactively classify all patches into fine-grained vulnerability types, even in the absence of prior CVE assignment.

Generally, patches addressing different categories of vulnerabilities exhibit distinct characteristics in their code modifications. 
Patches fixing memory corruption vulnerabilities typically involve memory management operations, such as adding bounds checks before memory accesses, nullifying pointers, or ensuring proper object lifetime management through careful allocation and deallocation.

Based on such observations, SID~\cite{wu2020precisely} is among the first to identify such patch patterns for OOB, UAF, and two other bug types, using pre-defined rules. The key idea is to leverage under-constrained symbolic execution to analyze the security impact before and after the patch. While the solution is highly precise—rarely producing false positives—its performance is greatly limited by its reliance on a highly restricted set of hard-coded patterns (e.g., only one for each bug type). Even from among supported cases, it misses 53\% of relevant patches due to the rigidity of its matching logic. Worse, patches that do not conform to any predefined pattern go undetected, leading to very high numbers of false negatives.  

Besides SID, there are several machine-learning-based solutions proposed to classify patches in recent years~\cite{pan2023fine,zhou2023colefunda}. They both attempt to classify patches into CWE labels, which include OOB and UAF bugs.
TreeVul~\cite{pan2023fine} leverages a pre-trained CodeBert~\cite{feng2020codebert} model and further fine-tunes the same for this task. Its input is the code diff in the patch, i.e., the added lines and removed lines, without any code context.
In comparison, CoLeFunDa~\cite{zhou2023colefunda} leverages contrastive learning and pre-trains a Bert model from scratch and fine-tunes it. Its input consists of program slices from the patched function.
Neither solution leverages the patch descriptions, i.e., commit titles and messages, for classification.
Both solutions also have only modest success. 
TreeVul's performance on Linux kernel patches is modest, with only a 66.98\% accuracy and an F1-score of 0.670. 
Similarly, CoLeFunDa achieved a precision of 0.52 an F1-score of 0.50, as reported in their paper.

Our work addresses this gap by introducing \system—a novel pipeline that combines large and small language models to achieve fine-grained vulnerability classification with significantly improved accuracy, \emph{even for patches without assigned CVEs or clear descriptions.}

\subsection{Machine Learning for Code Analysis}

Transformer-based architectures have been successful in NLP~\cite{vaswani2017attention,radford2018improving,radford2019language,brown2020language,devlin2018bert,lewis2019bart} and code understanding~\cite{feng2020codebert,guo2020graphcodebert,ahmad2021unified}.
Recent work has applied these models to a range of security applications.
For example, 
VulExplainer~\cite{fu2023vulexplainer} attempts to classify a given vulnerable function into associated CWEs (vulnerability types).
Sun \etal~\cite{sun2023silent} proposed an encoder-decoder framework that aims to detect and explain security patches (not explicitly focusing on classification).
Yu \etal~\cite{yu2020order} proposed hybrid models to perform binary code similarity detection.

The recent popularity of large language models~\cite{brown2020language, chen2021evaluating, chagptintro, openai2023gpt4, Llamaintro, Llama2intro, Llama31intro,openaio1}
has motivated researchers to investigate the use of LLMs for code related tasks.
They include, but are not limited to, patch generation~\cite{xia2023keep, pearce2022examining,ahmed2023better,alrashedy2024llmspatchsecurityissues}, fuzz testing~\cite{lemieux2023codamosa,yang2024kernelgptenhancedkernelfuzzing,yang2310white}, vulnerability detection~\cite{ullah2024llmsreliablyidentifyreason,wang2023defecthunternovelllmdrivenboostedconformerbased,sheng2024lprotectorllmdrivenvulnerabilitydetection}, bug reproduction~\cite{feng2023prompting}, and assisting static analysis~\cite{li2023hitchhikers,llift}.
While these studies demonstrate LLMs' potential across various code-related tasks, the specific challenge of identifying fine-grained vulnerability types such as OOB and UAF bugs from patches, remains unexplored. 
\section{Motivation and Overview}
\label{sec:method}

\subsection{Motivation}
\label{sec:context}

Our approach is motivated by two crucial observations from preliminary studies of security patch classification. 
Before detailing our complete methodology, we first discuss these critical observations, as they form the foundation of our technical approach and showcase the novelty of our contribution.
These observations not only reveal significant limitations in existing approaches but also guide our novel solution design.

\textbf{Observation 1: Untapped potential exists in commit descriptions.}

\begin{figure}[]
    \begin{tcolorbox}[boxsep=1pt, left=2pt, right=2pt, top=2pt, bottom=2pt]
    \small
    \textbf{ksmbd: fix use-after-free bug in smb2\_tree\_disconect} \\
    smb2_tree_disconnect() freed the struct ksmbd\_tree\_connect,but it left the dangling pointer. It can be accessed again under compound requests.
    \end{tcolorbox}
    \caption{Commit title and message of a Linux kernel patch with an explicit indication of the type of bug fixed}
    \label{fig:keyword-example}
\end{figure}
\begin{figure}[]
\begin{tcolorbox}[boxsep=1pt, left=2pt, right=2pt, top=2pt, bottom=2pt]
\small
{\textbf{ipv6: raw: Deduct extension header length in rawv6\_push\_pending\_frames}}

The total cork length created by ip6\_append\_data includes extension
headers, so we must exclude them when comparing them against the
IPV6\_CHECKSUM offset which does not include extension headers.
\end{tcolorbox}
    \caption{Commit title and message of a second Linux kernel patch with an implicit hint of the type of bug fixed}
    \label{fig:indirect-example}
    \vspace{-.2in}
\end{figure}
While it is known that the natural language descriptions in patches can reveal the nature of issues being fixed, they have been leveraged thus far for only coarse-grained patch classification (in the pre-LLM era), i.e., into security vs. non-security patches, like PatchRNN~\cite{wang2021patchrnn} and other works~\cite{zhou2021spi, zuo2023commit,zhou2017automated,sawadogo2020learningcatchsecuritypatches}.
State-of-the-art methods, like CoLeFunDa~\cite{zhou2023colefunda},TreeVul~\cite{pan2023fine} and SID~\cite{wu2020precisely}, for fine-grained patch classification, e.g., OOB vs. UAF bugs, however, \emph{completely overlook such valuable indicators.} 

We observe that patch commit descriptions can sometimes directly reveal the exact bug type with explicit indicators (e.g., ``use-after-free'') as shown in ~\autoref{fig:keyword-example}.
In other cases, the bug type is not directly mentioned, but can be inferred from the description (from a human perspective).
~\autoref{fig:indirect-example} illustrates such an example where a miscalculation of a length variable occurs. 
Although the bug type is not explicitly specified, the length variable adjustment is indicative of a potential out-of-bounds access vulnerability.
Surprisingly, our analysis of SID's dataset reveals the prevalence of such indicators: among the 227 security bugs that SID ultimately identified, our experiment shows that a remarkable \textbf{90.75\%} contain direct or indirect indicators of the vulnerability type in their commit titles or messages, suggesting the utility of patch descriptions.

\begin{figure}[]
    \begin{tcolorbox}[boxsep=1pt, left=2pt, right=2pt, top=2pt, bottom=2pt]
    \small
    { \textbf{Don't feed anything but regular iovec's to blk_rq_map_user_iov}} \\
    In theory we could map other things, but there's a reason that function is called "user_iov".  Using anything else (like splice can do) just confuses it.
    \end{tcolorbox}
    \caption{Commit title and message of one Linux kernel patch without hints of the type of bug fixed}
    \vspace{-.2in}
    \label{fig:noindication-example}
   
\end{figure}

\begin{figure*} 
    \begin{subfigure}[t]{0.48\textwidth} 
        \begin{lstlisting}[ breaklines=true]
@@ -316,6 +316,11 @@ int st21nfca_connectivity_event_received(struct nfc_hci_dev *hdev, u8 host,
 			return -ENOMEM;
 
 		transaction->aid_len = skb->data[1];

+
+		/* Checking if the length of the AID is valid */
+		if (transaction->aid_len > sizeof(transaction->aid))
+			return -EINVAL;
+

 		memcpy(transaction->aid, &skb->data[2],
 		       transaction->aid_len);


\end{lstlisting}
\caption{A patch for an out-of-bounds access vulnerability}
\label{fig:oob-example1}
 \end{subfigure}
    \hfill 
\begin{subfigure}[t]{0.48\textwidth} 
        \begin{lstlisting}[ breaklines=true]
@@ -567,12 +567,11 @@ static void nfsd_init_dirlist_pages(struct svc_rqst *rqstp,
 	struct xdr_stream *xdr = &resp->xdr;
 
-	count = clamp(count, (u32)(XDR_UNIT * 2), svc_max_payload(rqstp));
 	memset(buf, 0, sizeof(*buf));
 
 /* Reserve room for the NULL ptr & eof flag (-2 words) */
-	buf->buflen = count - XDR_UNIT * 2;
+	buf->buflen = clamp(count, (u32)(XDR_UNIT * 2), (u32)PAGE_SIZE);
+	buf->buflen -= XDR_UNIT * 2;
 	
    buf->pages = rqstp->rq_next_page;

    
\end{lstlisting}
        \caption{A second patch for an out-of-bounds access vulnerability}
        \label{fig:oob-example2}
    \end{subfigure}
    \caption{Examples of two common patch patterns for out-of-bounds access}
    \vspace{-.2in}
\end{figure*}

\textbf{Observation 2: Code diff patterns cannot effectively identify bug types in prior research.}
In the cases where no clear indicators are available in the patch description, e.g., ~\autoref{fig:noindication-example}  and ~\autoref{fig:diff}, code diff becomes a necessary signal for patch classification.
For example, an out-of-bounds (OOB) vulnerability is often patched by adding a bounds check (see ~\autoref{fig:oob-example1}), or by recalculating the size of critical memory areas (see ~\autoref{fig:oob-example2}).

However, some patches do not follow common patch patterns and thus require additional code context beyond immediate code changes (an example is shown in ~\autoref{fig:diff} discussed in detail later).
This reveals a fundamental challenge in patch classification: \emph{the true nature of a vulnerability fixed by a patch --- its root cause, impact, and security implications --- is often intertwined in intricate code relationships that extend far beyond the immediate patch vicinity.}
Consequently, correctly classifying such patches to their bugs requires precisely extracting the full bug-logic-relevant code context.

Previous approaches~\cite{wu2020precisely,pan2023fine,zhou2023colefunda} unfortunately have not developed an informed solution. 
SID~\cite{wu2020precisely} relies on symbolic rules (which are analyzed using symbolic execution) and support only a single patch pattern for each UAF and OOB vulnerability type. 
TreeVul~\cite{pan2023fine} uses only the added lines and removed lines as the input to a machine learning model, without any additional code context --- insufficient to identify bug types for some cases. 
Colefunda~\cite{zhou2023colefunda} does feed additional lines of code to a machine learning model via intra-procedural slicing. However, the standard slicing technique will capture irrelevant code context due to control dependence and lack of vulnerability context that may span function boundaries. We discuss these limitations in more detail in \S\ref{sec:sbart}.

{\em Summary:} The above observations reveal new opportunities to improve the classification of security patches, which directly motivate the design of our solution outlined in the next section.

\subsection{Solution Overview}
\label{sec:overview}

\begin{figure}
\centering
  \includegraphics[width=0.5\textwidth]{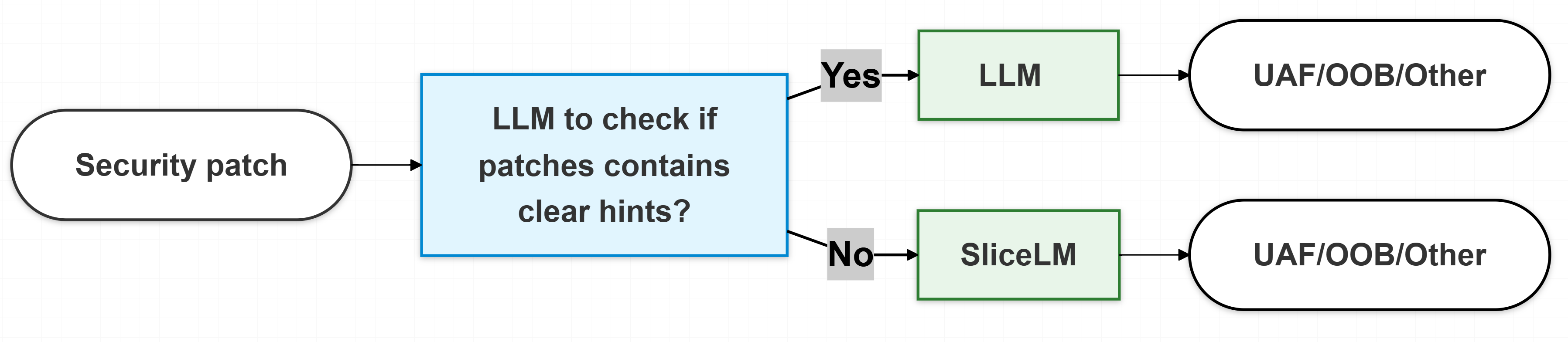}
  \caption{The high level pipeline of \system}
  \vspace{-.2in}
  \label{fig:pipeline}

\end{figure}

In this section, we present an overview of the key components of \system and highlight three core design strategies.

\PP{Design strategy 1: Classifying patches with an LLM.}
We argue that LLMs offer a superior solution for classifying patches compared to prior approaches. Recent advancements have demonstrated the exceptional ability of LLMs to handle both natural language and code-related tasks~\cite{zhao2023survey,openai2023gpt4,qin2023chatgpt,xia2023keep}. This makes them particularly well-suited for (1) analyzing patch descriptions, which often contain textual clues about bug types, (2) simultaneously processing the textual and code components of patches --- something static analysis struggles with, and
(3) new patch patterns can be relatively easily incorporated using the LLM's few-shot learning capabilities~\cite{brown2020language}, in contrast to rigid symbolic rules.

\PP{Design strategy 2: Classifying patches using a local model.}
While LLMs are very good at classifying patches with clear indicators,
our preliminary experiments find that the effectiveness decreases significantly when patches do not have sufficient hints.  
This motivates the collection of additional code context beyond the original code diffs in the patch. 
To this end, we developed a custom program slicing method that concisely captures the lines of code that constitute the bug, while minimizing the irrelevant lines (compared to the traditional slicing method).
Unfortunately, since our custom slices are in a new, specialized data format, we find that the LLM cannot effectively utilize these slices even when we attempt to teach them via in-context learning (as shown later in \S~\ref{subsubsec:ablation}). 
Thus, (given the unique nature of the data), we train a small language model, called \tool that can ingest such custom slices and infer the associated bug types.

\PP{Design strategy 3: A dual-method pipeline.} 
Given the above, we categorize patches into two classes: those that contain clear vulnerability indicators ({\em patches with hints}) and those that require more code-level analysis ({\em patches without hints}). This fundamental distinction motivates the design of \system, a dual-method pipeline that strategically handles each category with specialized techniques, as illustrated in ~\autoref{fig:pipeline}.
Specifically, we design our pipeline to first differentiate the two categories of patches leveraging an LLM.
Patches with hints are processed using methods that can effectively leverage their indicators, while patches without hints are analyzed by our dedicated model that focuses on careful examination of code modifications and the vulnerability context. In summary, as discussed below, an LLM excels at leveraging hints, but a dedicated model designed to understanding code context beyond code diffs is essential when such hints are absent.
This dual-method architecture stands in contrast to prior approaches like CoLeFunDa~\cite{zhou2023colefunda} and TreeVul~\cite{pan2023fine}, which rely on a single model throughout; by strategically integrating two models, \system more effectively leverages both high-level semantic cues and deep code context

\section{\system Design}
\label{sec:method}

In this section, we describe the two main components of \system: LLM-based patch classification and \tool (our own custom model). 
For the former, we focus on the design of prompt strategies that guide the LLM. For the latter, we focus on our custom slicing technique and how it is used to train our model.

Our approach assumes that each patch corresponds to a single vulnerability --- an assumption shared by prior work~\cite{pan2023fine,wu2020precisely,zhou2023colefunda}. This is further supported by the official Linux documentation~\cite{submitpatchrules}, which advises: ``Solve only one problem per patch.'', and is consistent with our empirical observations.

\subsection{LLM-based patch classification}
\label{sec:prompt}

We have two high-level goals when we leverage the LLM for our task. First, we aim to exploit an LLM to extract hints from patches that have them (Design strategy 1). Specifically, we look for hints in (1) the natural language description of commit titles and messages and (2) common patch patterns observed in commit diffs for bug types of interest, such as use-after-free (UAF) and out-of-bounds (OOB) access. Second, we seek to differentiate patches with and without hints, so that we can feed them to the corresponding methods in our pipeline (Design strategy 3).

In our design, we focus on guiding an LLM to effectively identify vulnerability indicators through both commit descriptions and common patch patterns. If successful, the bug type will also be reported and extracted subsequently. 
Otherwise, they are fed to our dedicated model.
LLMs are known to be extremely effective at processing natural language;
with respect to patch patterns, in contrast to SID~\cite{wu2020precisely} which uses precise but rigid symbolic rules to identify them, we hypothesize that LLMs can identify patch patterns in a more flexible way, e.g., recognizing small variations of the same underlying pattern.
To this end, we encode common patch patterns in the form of examples for few-shot learning~\cite{brown2020language}.

\PP{Patch pattern encoding (in-context learning).} The downside of not relying on precise symbolic rules is that it also creates room for misclassification, especially when ambiguities arise. To overcome this challenge, we present both UAF/OOB examples and non-UAF-OOB examples as  guidance for the LLM in the form of few-shot prompts~\cite{brown2020language}.
Specifically, our selected non-UAF-OOB examples cover various memory corruption bug types that involve changes to \textbf{critical memory operations} (e.g., initialization, free, and use).
Since UAF and OOB are both related to vulnerable memory operations, these chosen patches share more similarities with UAF and OOB compared to patches that do not involve vulnerable memory operations. \textbf{The overlaps of repair characteristics} in terms of vulnerable memory operations can cause confusion in a classification task. 
For example, patches addressing both UAF and memory leak bugs often involve memory deallocation operations, such as \texttt{free()}.
A memory leak bug is typically fixed by adding a \texttt{free()} somewhere. A UAF bug may be patched by deferring a \texttt{free()}~\cite{deferredfree} --- both may exhibit a line with added \texttt{free()}. 
These examples are not enforced as hard-coded decision rules (e.g., as in SID~\cite{wu2020precisely}). Instead, the LLM uses these examples inductively to generalize across diverse real-world patches. This enables \textbf{greater flexibility and adaptability}, especially when handling variations in patch style, naming, or structure. Thus, while \system incorporates human-defined examples, it avoids fixed symbolic rule logic, and relies instead on the LLM’s ability to reason and adapt.

By feeding explicit example patch patterns for these memory corruption bug types, we seek to help the LLM differentiate them better.
Specifically, the patterns we encode for \emph{different memory corruption bug types} are as follows:

\squishlist
\item {\em out-of-bounds access}: add boundary check; recalculate memory area sizes.

\item {\em use-after-free}: nullify pointers after freeing.

\item {\em null pointer dereference}: add null pointer validation checks.

\item {\em use before initialization}: add proper variable initialization. 

\item {\em memory leak}: insert memory free function calls along an execution path.
\squishend

\begin{figure}
\centering
  \includegraphics{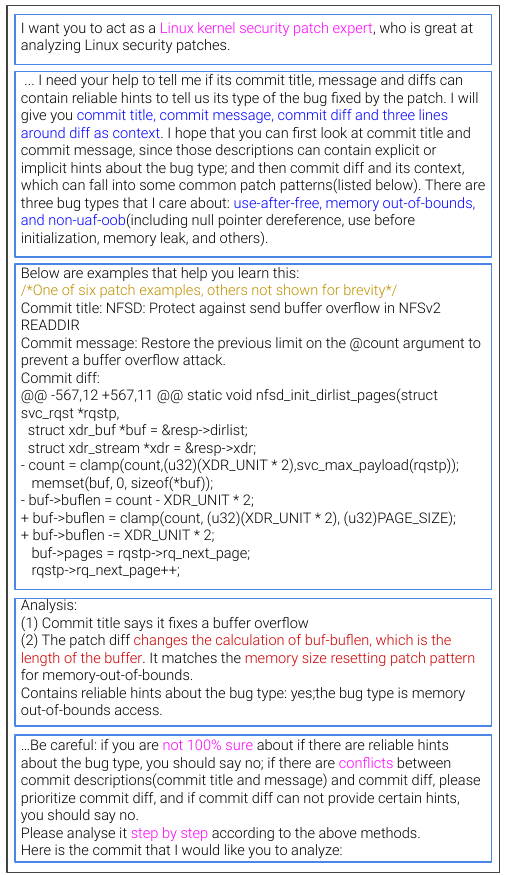}
  \caption{Structured prompt template for patch classification; the blue boxes delineate, in order, five key components: (1) role specification, (2) task description, (3) example, (4) analytical methodology, and (5) extra guidelines. }
\vspace{-.3in}
  \label{fig:prompt}
\end{figure}

For the above patch patterns, our prompts follow the chain-of-thought strategy~\cite{wei2022chain} by providing both representative patches and detailed explanations of how to interpret them and identify key indicators --- see the last two boxes in Figure~\ref{fig:prompt} which illustrates our prompt template. This approach guides an LLM through the intermediate reasoning steps needed to analyze patches, helping it decompose complex patch analysis tasks into manageable steps and ultimately improve its classification accuracy.

Note that our approach can be easily extended with more patch patterns by few-shot learning to provide examples in prompts.  
To demonstrate this, we expand the patch patterns for OOB vulnerabilities beyond what was supported by SID.
Specifically, while SID only accounts for the pattern of ``adding bounds checks'', our approach in addition incorporates ``recalculating memory area sizes'' as a supported pattern.
Extending the patterns and vulnerability types beyond what we have currently remains a direction for future work we can explore.

\begin{figure*}
\centering
  \includegraphics[width=0.8\textwidth]{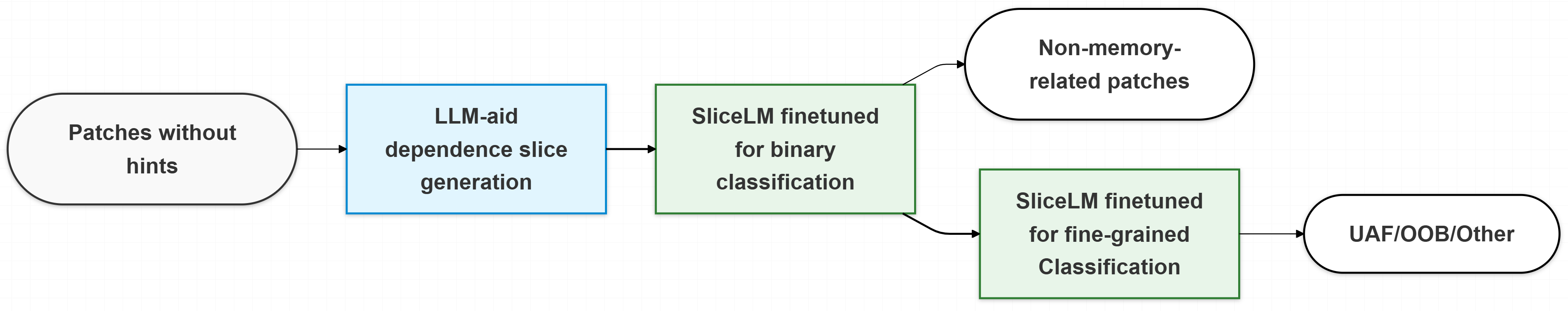}
  \caption{\tool's pipeline to classify \difficult
  }
  \label{fig:nohints}
\vspace{-.2in}
\end{figure*}
 
\PP{Prompt template}
In ~\autoref{fig:prompt}, we present the structured prompt template for LLM-based patch classification. As one can see, we structure the prompts into five parts. First, we establish the role specification, defining an LLM as a Linux kernel security patch expert. Second, we provide a detailed task description that outlines the analysis requirements, including the input format and the goal of the task. Third, we include an example (one of six patch patterns we support, others omitted for brevity) to demonstrate the expected analytical approach. Fourth, we present the analytical methodology, showing step-by-step reasoning that covers both natural language descriptions and code diff pattern interpretation. Finally, we  emphasize the importance of certainty in classification.

It is worth noting that towards the end of the prompt, we explicitly ask the LLM to ``give up'' on the classification if there are no reliable hints about the bug type. This is important because we do not want to force the LLM to make an inference unconditionally; otherwise, the results may be inaccurate.

\subsection{\tool-based patch classification}
\label{sec:sbart}

\tool is designed to classify \textbf{patches without hints}, i.e., patches that are classified by an LLM as such.
As discussed in \S\ref{sec:overview} (design strategy 2), proper code context is critical to identify the type of bug being fixed in a patch and the context needs to be carefully extracted to capture the true nature and critical elements of the vulnerability. Specifically, the context must be designed to \textbf{capture the lines of code most semantically related}. Accomplishing this is no easy task. Relying on the default context of three lines around code diffs risks overlooking critical patch-relevant details, while using entire patched functions or traditional program dependence slices can confuse the model with irrelevant lines of code. 

To solve this, we developed a custom slicing method.
Given our slices are specialized, we choose to pre-train a BERT-based language model from scratch (details will be presented in \S\ref{sec:impl}), to grasp dependence-based relationships between slices and patch diffs. This is different from prior work~\cite{pan2023fine} relying on the already pre-trained CodeBert model~\cite{feng2020codebert}. Then we fine-tune the model for two tasks: (1) the binary classification of memory corruption patches and other non-memory-related patches; and (2) the fine-grained classification of memory corruption patches into UAF, OOB and others. 
The workflow of SliceLM is presented in ~\autoref{fig:nohints}.

\subsubsection{Custom slicing}  
\label{sec:slicing}
Slicing is an appropriate solution to extracting the code context (\S\ref{sec:overview}). 
Ideally, slicing should include all relevant code --- but {\em only} the relevant code --- to capture the critical operations tied to the bug's behavior.
For example, 
we would like to find the lines of code that indicate a freed object being \emph{used} subsequently, for a UAF bug. However, it is hard to determine how far away such operations are from the changed code lines in the code diff.
If we arbitrarily increase the scope of a slice, e.g., making it inter-procedural, we would encounter significant noise from irrelevant code inclusion and face scalability challenges.
To address this challenge, we develop three novel strategies: \emph{(1) a selective slicing heuristic to prioritize data dependence over control dependence,  (2) an LLM-aided function renaming method to enhance the vulnerability context while avoiding inter-procedural analysis; (3) an LLM-aided strategy to prune redundant code changes.}
We detail the design of the three strategies next.

\PP{Selective slicing rule (avoiding bloating)}
We take the removed and added code of a patch as the slicing criterion, i.e., the starting point of slicing~\cite{weiser1984program}, because the variables changed in these lines are relevant to the bug.
We observe that traditional program slicing often produces excessively large slices by including all data and control dependencies—even those unrelated to the bug, such as variables and operations that do not impact its presence or absence.

\begin{figure}[h]
\centering
  \includegraphics[width=0.5\textwidth]{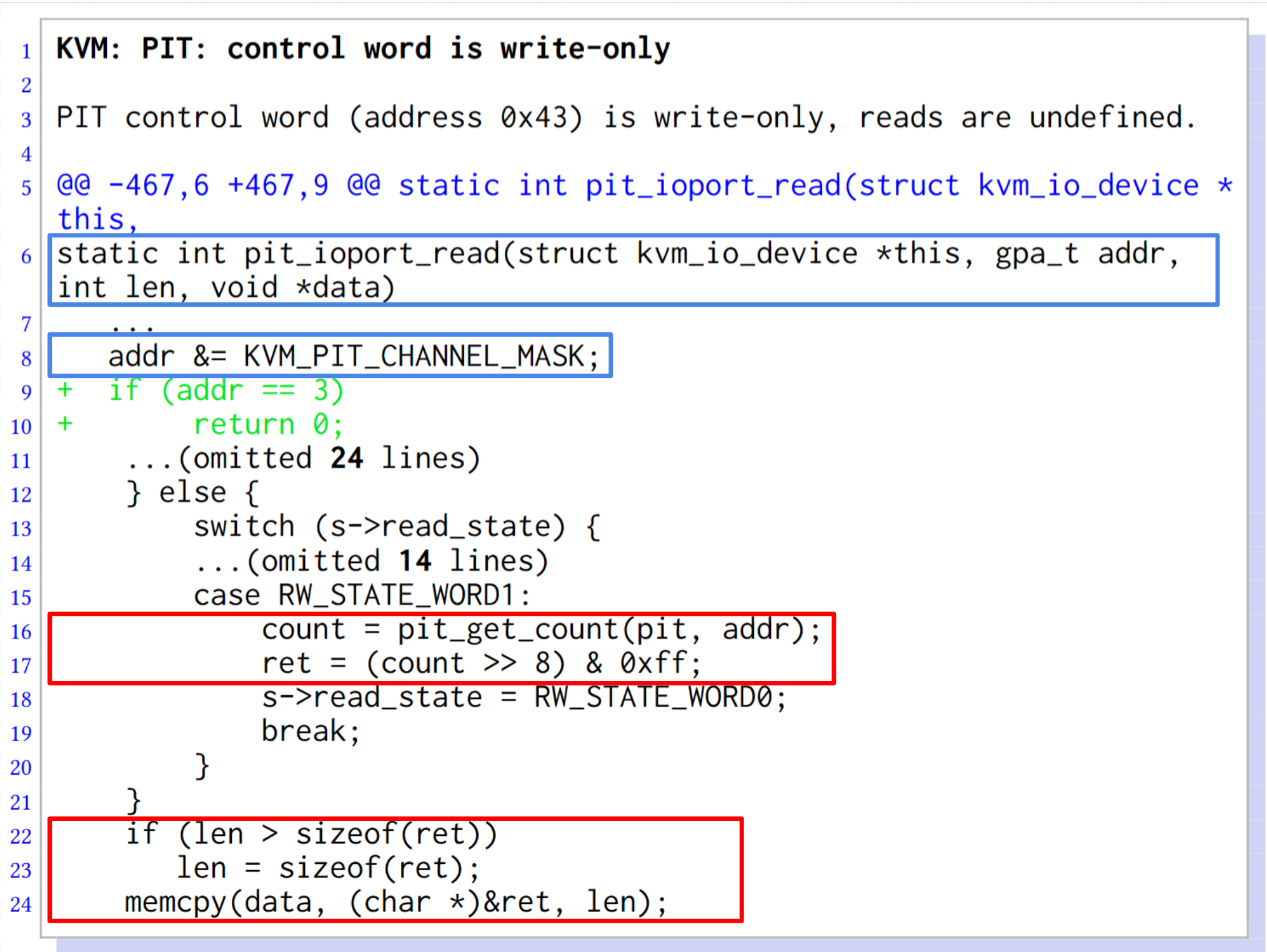} 
  \caption{A simplified patch for memory out-of-bounds vulnerability (\textbf{Blue} boxes indicate the extracted backward slices and  \textbf{red} boxes indicate the extracted forward slices using our method)}
  \label{fig:diff}
\vspace{-.2in}
\end{figure}

In particular, control dependence can lead to bloated slices.
To illustrate, consider the example patch in \autoref{fig:diff}. The patch has introduced two additional lines of code that would return early  if a specific condition is met. 
If we apply the standard forward control dependence slicing on the conditional statement, the slice would include all the lines after line 10. This is because whether these subsequent lines will be executed is control-dependent on the conditional statement, i.e., lines after 10 will not be executed if the new condition on line 9 is true.
However, we know that the majority of the subsequent lines of code are not relevant to the bug. Instead, only the key operations directly involving the variable being checked (in this example \texttt{addr}) should be retained in the slice.

To mitigate such control-dependence-induced bloating, we redefine the handling of control dependence by introducing a focused, variable-driven slicing rule that balances precision and relevance.
Instead of including all control-dependent statements, our approach selectively incorporates statements based on their direct interaction with the slicing criterion:
\emph{(1) forward control dependence slices include only statements that use variables involved in the criterion (e.g., conditional checks), and 
(2) backward control dependence slices include only conditional statements that use variables in the criterion.
This hybrid approach blends aspects of control dependence and data dependence slicing, prioritizing patch-relevant impact while minimizing irrelevant code.}
The intuition is to limit the inclusion of code due to control dependence to those that are more likely to contribute to understanding the impact of patches, e.g., checking a length variable or the validity of a pointer.

\phantomsection
\label{ex:myexample}

To illustrate our approach, let us look at the slices we obtain if we use line 9 of ~\autoref{fig:diff} as the slicing criterion (since line 9 is an added line). 
Backward slicing yields lines 6 and 8 due to data dependence for variable \texttt{addr}. 
Traditional forward slicing will include all of the code after line 10 (more than 40 lines), since their execution will be controlled by the if-return check in lines 9 and 10.
However, as discussed previously, our selective slicing will include only the lines which use the variable checked in the if condition; therefore, line 16 is included in the forward slice (which is directly related to patch semantics); line 17 is included since it uses the variable \texttt{count} defined in line 16; line 20, 21 and 22 are also included due to the use of variable \texttt{ret} defined in line 17.
In other words, our forward slicing yields lines 16, 17, 22, 23, and 24, which captures all operations relevant to the OOB vulnerability and avoids significant noise compared to the 40+ lines present in the forward slice of the traditional method. 
Using the slice, we can see that the patch adds a check on a variable \texttt{addr}. This variable \texttt{addr} is an index that is used to retrieve an address \texttt{ret} and the size of source data to be copied. Without imposing constraints on 
\texttt{addr} (line 9-10), it is possible that the size of the source data is bigger than that of the destination data during \texttt{memcpy()}, causing a memory out-of-bounds access in line 24. 
These lines are necessary and sufficient for the ML model to make an inference that an OOB vulnerability is  being fixed by the patch.
If the traditional program slicing method is used, we observe that our model will misclassify this case.

\PP{LLM-aided slicing (function renaming).}\label{ref:llmaid}
As shown in the example in \autoref{fig:diff}, obtaining complete vulnerability context often requires tracking operations that are distant from the code diff, e.g., \texttt{memcpy()}. 
In fact, it can even require tracking across function boundaries through inter-procedural analysis. 
However, performing inter-procedural slicing can be prohibitively expensive and may introduce substantial amounts of irrelevant code, potentially degrading the performance of the model we aim to train. Therefore, we opt for intra-procedural slicing, though this inevitably excludes relevant context from other functions.

To address this limitation, we leverage a key observation that function semantics are often revealed in function names: they typically provide clear indicators of their behaviors~\cite{linuxcoding}. For example, the function \texttt{drm_mm_remove_node()} suggests its purpose of node removal. 
Since function names are included in the code diffs and slices, they should in principle already contain semantics that can benefit patch classification. However, the sheer diversity of functions and their names can be a hurdle in training our local model: many projects define wrappers or custom-named functions for memory operations like allocation, freeing, reference counting, or copy operations, using names that are not standardized or intuitive.

We leverage an LLM to interpret and analyze the functionality of open-source functions (e.g., in the Linux kernel) and convert them into \textbf{more standardized function names} to aid the training and classification process. This capability stems from LLMs being pre-trained on extensive datasets (including Linux source code and documentation), a finding also confirmed by prior work~\cite{llift}.
Specifically, we rename functions according to their relevant \emph{memory operations} because these behaviors are closely tied to memory corruption vulnerabilities. Our approach asks an LLM to \textbf{identify and rename functions based on key memory operations} including allocate, free, read, write, map, copy, and reference count increase/decrease. When an LLM identifies that a function's primary functionality is one of these operations, it renames the function accordingly; otherwise, the original name is preserved. For instance, \texttt{l2cap_chan_hold_unless_zero()} is renamed to \texttt{increase_reference_count_if_not_zero()} to explicitly reflect its reference count operation.
Finally, we replace original function names in custom slices and code diffs with their semantically renamed counterparts. As more patches are analyzed, we accumulate these mappings and cache them to minimize inference overhead. This enables the downstream model to reason about what a function does, rather than what it is called.

Standardizing function names semantically creates a unified vocabulary for memory operations, enabling our model to more effectively identify and correlate similar memory behaviors across different code segments.
Considering the example in ~\autoref{fig:funcrename},
the UAF issue arises because \texttt{put_cred(creds)} is executed even if the variable \texttt{ret} is `1' (which is a valid return value), causing an unwanted reference count decrement and potentially triggering an unwanted free.
Our function renaming approach assigned the name \texttt{decrease_credential_reference_count()} to the function \texttt{put_cred()}, emphasizing its connection to the reference counter decrement.
Without this renaming, we found that our model would misclassify the case as OOB instead.

\begin{figure}[t]
    \centering
    \begin{minipage}[t]{1\linewidth}
\begin{lstlisting}[]

We currently check for ret != 0 to indicate error, but '1' is a valid return and just indicates that the allocation succeeded with a wrap. Correct the check to be for < 0, like it was before the xarray conversion.

@@ -9843,10 +9843,11 @@ static int io_register_personality(struct io_ring_ctx *ctx)
 	ret = xa_alloc_cyclic(&ctx->personalities, &id, (void *)creds,
 			XA_LIMIT(0, USHRT_MAX), &ctx->pers_next, GFP_KERNEL);
-	if (!ret)
-		return id;
-	put_cred(creds);
-	return ret;
+	if (ret < 0) {
+		put_cred(creds);
+		return ret;
+	}
+	return id;
 }

\end{lstlisting}
    \end{minipage}
    \caption{An example where function renaming  helps}
    \label{fig:funcrename}
    \vspace{-.2in}
\end{figure}

\PP{LLM-aided slicing (pruning redundant code changes)}
The above method is sufficient when we consider relatively small patches. However, large patches in the real world can contain numerous code changes spread across multiple functions and even files; 
\texttt{yet, not all changes target fixing the vulnerability.}
Unfortunately, we cannot directly ask an LLM to pinpoint the bug-logic-relevant changes. This is because the patches we process at this stage are already considered ``without hints'', and pinpointing the bug-logic-relevant changes would require clues about the bug type. 

That said, there are still opportunities to prune clearly redundant code changes in a bug-agnostic way.
Specifically, we propose to leverage an LLM to identify two patterns of redundant code changes:
\textit{(1) redundant changes (same code changes that appear in multiple places due to the change of function call parameters), and (2) code refactoring without changing the actual code behaviors.}

\begin{figure}[h]
\centering
  \includegraphics[width=0.5\textwidth]{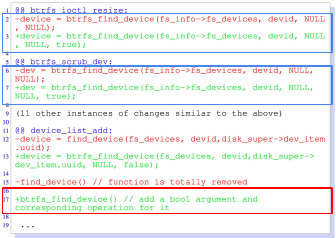} 
  \caption{A simplified patch for redundant changes example; Blue boxes are semantically equivalent, red boxes are critical lines}
  \label{fig:truncate}
\vspace{-.2in}
\end{figure}

\begin{figure}[h]
\centering
  \includegraphics[width=0.5\textwidth]{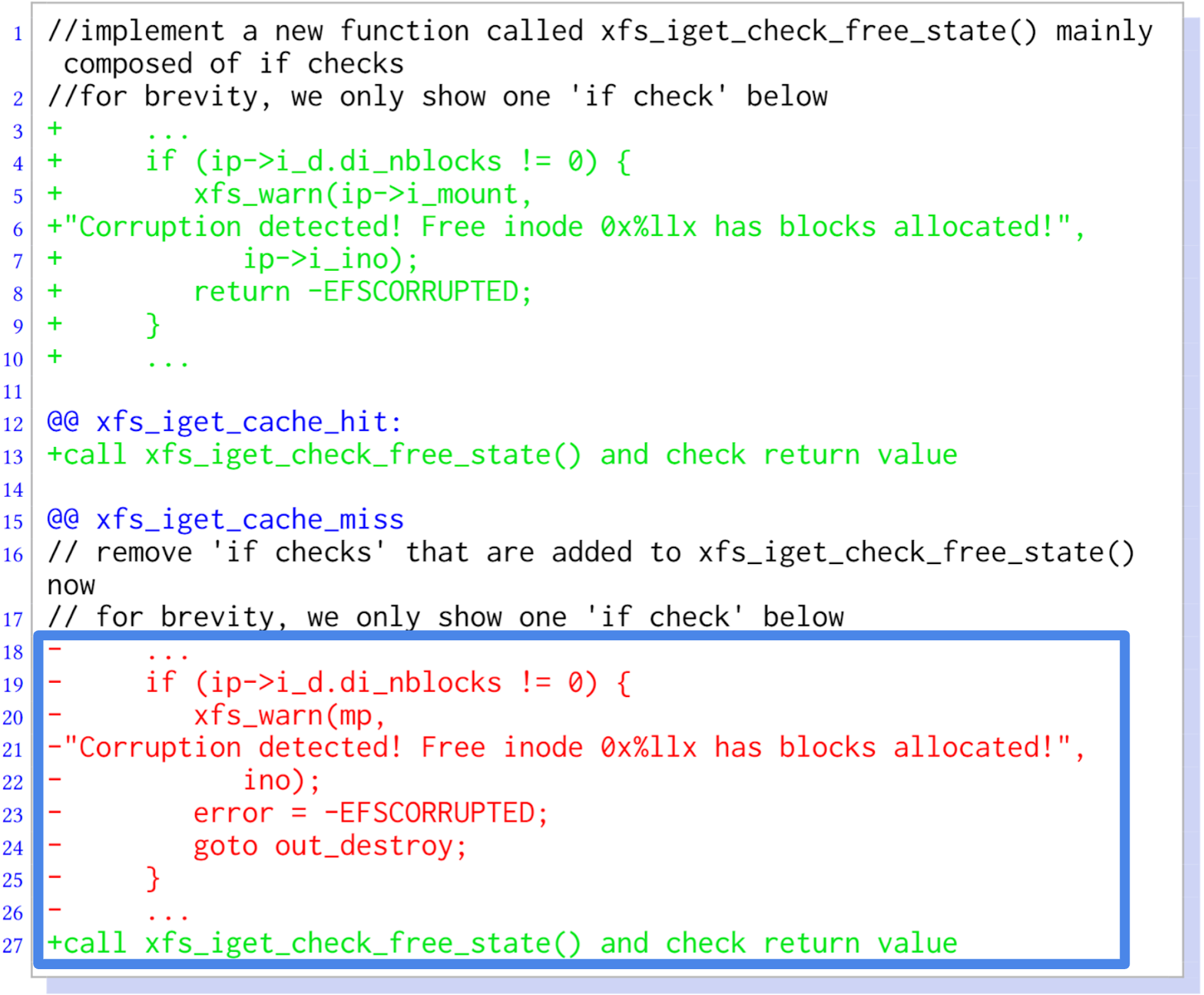} 
  \caption{A simplified patch for code refactoring example}
  \vspace{-.2in}
  \label{fig:truncate2}
\end{figure}

~\autoref{fig:truncate} illustrates an example with many change sites that are basically redundant. The patch is composed of \textbf{17} change sites (for brevity, we only show 6). However, the first \textbf{13} change sites are semantically equivalent, since all of them are adding a \texttt{true} argument when calling \texttt{btrfs_find_device()},  as shown in the first two change sites highlighted in the two blue boxes in ~\autoref{fig:truncate}.
We find that the critical lines for the patch are unrelated to {\em any} of repeated instances, e.g., the code within blue boxes in  ~\autoref{fig:truncate} that is repeated.
Including all these ``similar'' changes and their corresponding dependence-based context would not only obscure the core vulnerability-fixing modifications (the meaningful change is highlighted in the red box) 
but also consume valuable space in our
model's limited input length (1,024 tokens). The LLM provides us with an effective means of identifying such cases and preserving only a single representative instance of multiple equivalent code modifications.

~\autoref{fig:truncate2} 
depicts a simplified example where the changes highlighted in the blue box represent a modification that does not alter the underlying semantics of the code --- it merely relocates ``if'' checks into a new function call. 
For such refactoring, we ask an LLM to identify and remove such added and deleted code segments. 
After pruning, it is clear that the patch in \autoref{fig:truncate2} mainly added additional checks for \texttt{xfs_iget_cache_hit()}, as shown in the second change site.
This selective preservation produces a concise input for our model (\tool) by retaining only essential code modifications and discarding irrelevant or redundant ones.

\subsubsection{Two-stage classification for \difficult}

We fine-tune the pre-trained model for two downstream tasks, which are two classifiers, as shown in ~\autoref{fig:nohints}.
\emph{The first classifier is a binary one, classifying the patches into memory corruption vs non-memory-related.}
As noted in~\S\ref{sec:prompt}, the memory-related bug types we consider include: use-after-free, memory out-of-bounds access, use-before-initialization, null pointer dereference, and memory leak.
This is motivated by the observation that memory corruption patches inherently exhibit distinct characteristics from other types -- they tend to involve memory operations such as \texttt{free()} and \texttt{memcpy()}. Successfully ruling out non-memory-related patches can help improve the accuracy of the subsequent classifier (as will be shown in \S~\ref{sec:two-stage}).
\emph{The second classifier then distinguishes the patches into UAF, OOB, and other memory-corruption patches.}

\section{Implementation \& Experiment setup}
\label{sec:impl}

\subsection{Implementation Details}

\noindent \textbf{LLM usage.} Our experiments were conducted primarily using OpenAI GPT-4-turbo~\cite{gpt4turbo}.
In addition, we perform experiments using the Llama 3.1 405B~\cite{Llama31intro} and OpenAI-O1~\cite{openaio1}.

\PP{Custom slicing.}
We implemented our custom slicing at the source code level, using Joern~\cite{yamaguchi2014modeling,joerdoc}. 
The slice is computed for each change site, which includes the added lines and/or removed lines (with the plus and minus signs preserved) as well as the forward and backward slices. 

\PP{Additional information besides slices.}
In addition to the custom slices, we also feed a single-sentence summary of the patch description to our model. The summary is generated by leveraging LLM's summarization capabilities to distill verbose commit messages into a concise sentence.
The reasoning is that commit messages often contain useful information, e.g., about how a variable is used. Even if it does not directly hint at the bug type, it can potentially provide additional information when combined with slices.
Nevertheless, feeding long commit messages to our local language model consumes significant token space, making summarization necessary.

\subsection{Experiment Setup}
\label{sec:setup}

The pre-training of the model typically took approximately six days.
On average, the classification of a test case was achieved within 30 seconds (via the entire pipeline). Further details about model training (e.g., model architecture, optimization settings) are provided in Appendix~\ref{sec:modeltrain}.

\PP{Pre-training dataset.}
\tool is pre-trained on a comprehensive dataset of all available historical commits since the inception of the Linux kernel repository up to kernel v6.0, encompassing 1.1M commits. This process enhances the model's capability to grasp dependence-based relationships between slices and patch diffs, thereby capturing semantic features within the source code. We subsequently perform two separate fine-tuning processes on the pre-trained model: one for binary classification and the other for multi-class classification.

\PP{Fine-tuning dataset.} \label{ref:finetunedataset}
We prepare two datasets, one for each of the two fine-tuned tasks. For the multi-class classification, we select patches with standardized phrases in commit titles that indicate the type of the fixed bug to collect ground truth. For example, phrases like ``Fix use-after-free" or ``Fix out-of-bounds" suggest the type of the fixed vulnerabilities. We compiled \textbf{a list of common key phrases} (show in Appendix~\ref{sec:keyphrases}) used in the Linux kernel community and mapped them to the three bug types. This key-phrase-based labeling strategy generated a dataset of 10,540 samples.

To create the memory corruption and non-memory-related datasets, we used the above samples as positive samples (related to vulnerable memory operations) and commits labeled as ``non-security" in \texttt{PatchDB}~\cite{wang2021patchdb} as negative samples. It is worth noting that finding non-memory-related security patches is challenging, as there are many types of vulnerabilities which are difficult to enumerate. We believe the non-security patches in \texttt{PatchDB} provide a reasonable approximation of non-memory-related security patches, since non-security patches are also technically free of any vulnerable memory operations.

\PP{Evaluation dataset \#1: CVE patches}
Our first evaluation is on a quality-controlled dataset of Linux kernel CVE patches.
Since CVE patches are, by definition, confirmed security patches, we do not need to apply a binary classifier to distinguish between security and non-security patches. Instead, we can directly evaluate fine-grained security patch classification methods, e.g., such as \system and TreeVul. This allows a fair comparison, as all evaluated methods operate on the same set of established security patches.
Please note that this is akin to taking the output of an “ideal binary classifier” and feeding the same as the input to a fine-grained classifier (the latter being the focus of our effort). If a non-ideal binary classifier is used (wherein it does not yield a 100\% accuracy in classification results either due to improper training sets, overfitting or otherwise), the results will negatively influence the performance of the overall pipeline but is not directly attributable to the performance of the fine-grained classifier.
This collected CVE dataset consists of 946 patches manually assigned CWE labels, from 2015 to May 2024.
They include 78 CVEs published in 2024, which is after the knowledge cutoff date of the LLM we used in the evaluation. This can help us understand the potential data leakage problem where an LLM might gain an advantage because it might already have learned the details about the older CVEs in training data. 
Note that if evaluated cases occur in the fine-tuning dataset, they are excluded from the dataset during training to prevent overlap. 
According to CWE labels, we mapped these CVE patches to OOB, UAF, and non-UAF-OOB, by manually verifying the labels. More details of the data cleaning can be found in Appendix~\ref{sec:clean}).
Since this dataset is \textbf{quality-controlled and complete on the labels,} the goal of the evaluation is the comprehensive evaluation of classification performance, 
enabling both a systematic comparison with state-of-the-art approaches and detailed ablation studies.
\emph{We present the results of this evaluation in in \S~\ref{sec:cveeval}.}

\PP{Evaluation dataset \#2: Unlabeled Patches}
Second (and perhaps more critically), we challenged DualLM with 5,140
Linux kernel patches to assess its ability to discover previously unknown UAF and OOB bugs --- a test that mirrors the actual deployment scenarios where automated patch classification can prevent potential security breaches. 
Note that \system is designed to operate on security patches as input (same with TreeVul). Therefore, to evaluate its performance in a realistic end-to-end setting—where the security relevance of a patch may not be known in advance—we combined it with various upstream binary classifiers (security vs. non-security). This enables an assessment of how well the full pipeline performs in practice, from initial security identification to fine-grained classification. Differently, SID does not require the input to be security patches.
These patches were selected randomly from a total of 12K patches on the Linux kernel Long-Term Support (LTS) version 6.6 branch, as long as their patch date was in 2024 (after the knowledge cutoff date of the LLM we used). We choose these patches regardless of whether they are assigned CVE labels.
\emph{The results of this evaluation are presented in \S~\ref{sec:randomeval}.}

\PP{Comparison setup.}\label{sec:comparisonsetup}
We choose TreeVul~\cite{pan2023fine} and SID~\cite{wu2020precisely} as comparison targets since they are the state-of-the-art open-source solutions that share similar goals with our work. For TreeVul, since it classifies patches into CWE labels (including types beyond OOB and UAF), we mapped these labels to our three target categories: OOB, UAF, and non-UAF-OOB, consistent with our classification scheme. Furthermore, to ensure fairness and avoid data leakage, we excluded the CVE patches present in TreeVul's original training dataset from our CVE-based evaluation dataset. Note that we do not directly compare with CoLeFunDa~\cite{zhou2023colefunda}, because we are unable to obtain its source code.

For SID, which employs a rule-based matching method to specifically identify use-after-free, out-of-bounds, null pointer dereference, and use-before-initialization bugs, there is no explicit category corresponding directly to our non-UAF-OOB classification. To align SID's outputs with our categories, we adopted the following approach: 1) patches matched by SID are classified into their corresponding bug types (UAF or OOB), and 2) unmatched patches, as well as patches matched as null pointer dereference or use-before-initialization, are grouped into non-UAF-OOB. Since SID uses rule-based matching rather than machine learning, it does not require training data; therefore, our evaluation dataset for SID includes the entire CVE dataset without exclusions.

\begin{figure}[h]
\centering
  \includegraphics[width=0.5\textwidth]{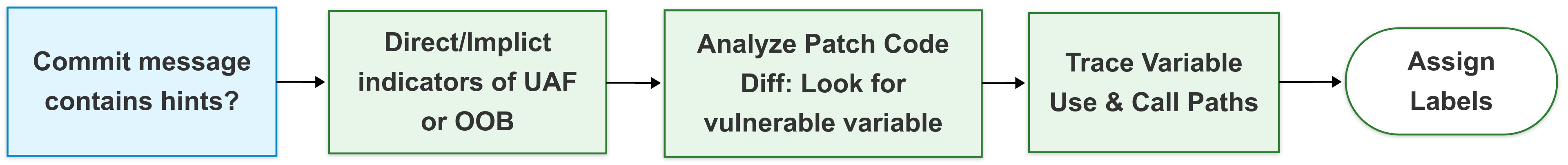}
  \caption{The manual verification process for real world patches}
  \label{fig:manualverifcation}
\vspace{-.2in}
\end{figure}

\PP{Manual verification.}\label{sec:manualverification}
To support the evaluation of our 5,140 real-world patches in terms of ground-truth validation, we perform targeted manual verification, as described below.
Labeling all 5,140 patches is an extremely time-consuming and labor-intensive task, making full manual annotation infeasible. To balance rigor and scalability, we adopt a tiered verification strategy:
(A) We manually verify all UAF and OOB cases identified by our pipeline using a structured multi-step process grounded in kernel security expertise (described below and illustrated in~\autoref{fig:manualverifcation});
(B) For patches classified as non-UAF-OOB, we sample 50 cases for manual review to estimate the false negative rate;
(C) We select representative cases to develop proof-of-concepts (PoCs) and real exploits to further validate classification correctness.
With respect to (A), our multi-step process is as follows: 1) \textbf{Commit Message Inspection}: We first check whether the commit title/message contains explicit or implicit hints. Phrases like “fix use-after-free” are considered reliable indicators, as kernel developers generally document key issues therein. 2) \textbf{Locate Vulnerable Variable}: We examine code changes. With respect to OOB, we look for signals such as bounds checks, size clamping, array index corrections, in order to locate the possible vulnerable variable. Next, we manually trace across function boundaries whether the modified variables are involved in memory accesses, such as indexing or \texttt{memcpy()} operations. With respect to UAF, we identify signals such as pointer nullifications after free, reference count adjustments, introduction of locks, or reordering of deallocation logic, to locate the possible vulnerable variable. We also review whether a function indirectly performs a free (via wrappers), and whether the code adds or delays such a free. 3) \textbf{Reconstruct the Vulnerability Logic}: For complex or ambiguous cases, we reconstruct the calling context based on clues in the commit message or associated stack traces (if available). We trace data flows and call chains across functions to identify the vulnerable logic path and verify whether the bug could lead to memory corruption.
4) \textbf{Conservative Labeling}: We assign labels only when confident. Uncertain cases are categorized as non-UAF-OOB to ensure dataset integrity. 
Five analysts with security expertise independently follow this procedure. In cases of disagreement, the analysts discuss the patch collectively and reach a consensus through majority agreement. 
On average, the manual analysis of a single case by one analyst takes approximately 20 minutes, highlighting the depth of inspection involved.
 
\PP{Metrics.}
For most experiments, we report accuracy, precision, F1-score, false positive rate, and false negative rate.
For multi-class classification evaluations, metrics such as precision can be more challenging to interpret, as multiple types of errors (e.g., predicting class A when it is actually class B, C, or D) must be accounted for.
Therefore, we use the commonly adopted \textit{weighted averaging} method~\cite{bishop2006pattern}, to assess the overall model quality in performing multi-class classification tasks.
The method essentially calculates metrics like \textit{precision}, and \textit{F1-score} individually for each class, and subsequently aggregates a weighted average by assigning weights to all classes according to their frequencies.

\section{Experimental Results}
\label{sec:eval}

\subsection{Evaluation on Quality-Controlled CVEs}
\label{sec:cveeval}

\subsubsection{Main Results}

\autoref{table:performance} demonstrates \system's effectiveness, with an overall \textbf{87.4\%} accuracy, \textbf{87.7\%} precision, and a \textbf{0.875} F1-score, while maintaining a low false positive rate of \textbf{5.4\%} and a relatively low false negative rate of \textbf{10.0\%}. 
Since our solution is a dual-method pipeline, we also break the results down by method. We find that 77\% of the evaluated patches are classified as \easy\ and 23\% are classified as \difficult. For patches considered
with hints or indicators, \system achieves a 90.3\% accuracy with a 90.7\% precision and a low false positive rate of 4.0\%. Even for those without indicators, our two-stage classification approach maintains high effectiveness, achieving an overall 77.9\% accuracy. Note that although the binary classification and fine-grained classification achieved 84.8\% accuracy and 81.8\% accuracy respectively, their combined accuracy is lower because errors from the two stages compound when the results are composed.
In summary, we believe these results are significant given the challenging nature of patch classification, especially when compared against the state-of-the-art, which we describe next.

\begin{table*}[]
\centering
{
\begin{tabular}{|l|c|c|c|c|c|}
\hline
\multicolumn{1}{|c|}{Task}  & Accuracy & Precision & F1-score & FP Rate & FN Rate \\
\hline
\textbf{Overall pipeline} & \textbf{87.4\%} & 87.7\% & 0.875 & 5.4\% & 10.0\% \\
\hline
\quad $-$ \textbf{Classification on patches with hints} & \textbf{90.3\%} & 90.7\% & 0.904 & 4.0\% & 6.6\% \\
\hline
\quad $-$ \textbf{Classification on patches without hints} & \textbf{77.9\%}  & 78.4\%   & 0.776    & 13.4\% & 24.1\% \\
\hline
\qquad $|$ Binary classification & 84.8\% & 85.8\% & 0.847 & 13.9\% & 16.5\% \\
\hline
\qquad $|$ Fine-grained classification & 81.8\% & 83.5\% & 0.821 & 8.8\% & 18.8\% \\
\hline

\end{tabular}
}
    \caption{The performance of \system }
    \vspace{-.2in}
\label{table:performance}
\end{table*}

\subsubsection{Comparative study}

To further validate \system's effectiveness, we compare its performance against that of two open-source, state-of-the-art solutions, viz.,  TreeVul~\cite{pan2023fine} and SID~\cite{wu2020precisely}. 
The comparison setups are described in \S\ref{sec:manualverification}.

In \autoref{table:treevul}, we show the comparison with TreeVul and see a substantial performance gap: TreeVul achieved only a \textbf{65.6\%} accuracy with an F1-score of 0.653, while \system achieved significantly better performance with a \textbf{86.8\%} accuracy and an F1-score of 0.869.
Upon examining some specific cases, we identify two major reasons for the performance discrepancy:
(1) misclassified cases by TreeVul actually have indicators in the commit titles and/or messages, which allows \system to correctly classify them;
(2) misclassified cases by TreeVul needs the proper context to infer the correct type. TreeVul only takes code diffs as input, ignoring the commit title/message and code context.

 \begin{table}[]
\centering
{
\begin{tabular}{|rl|l|l|l|l|l|l|}
\hline
\multicolumn{2}{|r|}{Method}  & Accuracy & Precision & F1-score & FP rate & FN rate \\ \hline
\multicolumn{2}{|r|}{TreeVul} & 65.6\%   & 72.5\%    & 0.653    & 21.6\%   & 23.0\%  \\ \hline
\multicolumn{2}{|r|}{\system}  & 86.8\%   & 87.5\%    & 0.869    & 5.1\%   & 9.3\%  \\ \hline
\end{tabular}
}
\caption{Comparison between TreeVul and \system.To ensure a fair comparison and prevent data leakage, we excluded the CVE patches used in TreeVul's original training dataset}
    \vspace{-.1in}
\label{table:treevul}
\end{table}

~\autoref{fig:funcrename} illustrates an interesting example where TreeVul failed to identify the bug type. 
As discussed in \S~\ref{ref:llmaid},
our function renaming, which gave the new name \texttt{decrease_credential_reference_count()} for \texttt{put_cred()}, plays an important role in the correct identification of the associated UAF bug.
However, TreeVul does not appear to capture the knowledge about what \texttt{put_cred()} does (perhaps due to limited training data). In addition, we suspect that TreeVul's automatic alignment mechanism actually makes it pay less attention to \texttt{put_cred(creds)} because the function is present in both the pre-patch and post-patch version. 
Additional case studies are presented in Appendix~\ref{subsec:treevulcases}.

\begin{table}[]
\centering
{
\begin{tabular}{|rl|l|l|l|l|l|l|}
\hline
\multicolumn{2}{|r|}{Method}  & Accuracy & Precision & F1-score & FP rate & FN rate \\ \hline
\multicolumn{2}{|r|}{SID} & 63.8\%   & 77.7\%    & 0.546    & 21.2\%   & 67.8\%  \\ \hline
\multicolumn{2}{|r|}{\system}  & 87.4\%   & 87.7\%    & 0.875    & 5.4\%   & 10.0\%  \\ \hline
\end{tabular} 
}
\caption{Comparison between SID and \system}
\vspace{-.2in}
\label{table:sid}
\end{table}

~\autoref{table:sid} shows the comparison with SID, and the results demonstrate a similar performance gap: SID achieves an accuracy of \textbf{63.8\%} and an F1-score of 0.546, compared to \system's \textbf{87.4\%} and 0.875, respectively. Importantly, SID’s false negative rate is alarmingly high at 67.8\%, meaning that it misses more than half of the vulnerabilities. In contrast, \system achieves a much lower false negative rate of 10.0\% while also reducing the false positive rate to 5.4\% (compared to SID’s 21.2\%).\footnote{SID’s performance on our evaluation dataset appears significantly lower than what was reported in the original paper. We consulted the SID authors when we ran SID against the evaluation dataset to ensure that we ran it correctly. One possible reason is the difference between the two evaluation datasets. The 97\% precision reported in the SID paper is on a different dataset than ours (we used the CVE dataset, including many patches that were published after the SID paper was published).}

One might question whether \system’s use of multiple few-shot examples, including those outside the scope of SID (e.g., ``memory area size recalculations”), gives it an unfair advantage. To address this, we repeated the evaluation with a constrained version of \system using only one OOB patch example that matches the patch pattern used by SID (i.e., boundary check additions). 
The result only degraded slightly and remains strong: \system achieves an accuracy of 86.9\% and an F1-score of 0.869, still significantly outperforming SID.

Our analysis shows that SID's effectiveness is limited by its reliance on symbolic pattern matching rules, which support only one pattern per vulnerability type. For instance, in detecting out-of-bounds vulnerabilities, SID looks exclusively for boundary check additions, missing other common fixes like memory area size adjustments or more nuanced protection mechanisms (as illustrated in ~\autoref{fig:diff}).

\subsubsection{Generalization to other bug types and beyond Linux}
Although we have focused on the most severe types of bugs, i.e., UAF and OOB, we have also evaluated \system's generalizability by training/testing it to classify other bug types. 
For this expanded experiment, the binary classification step (`memory corruption vs. non-memory') requires no retraining. However, the fine-grained classification step does require retraining. Specifically, we reuse the same pretrained model and apply the same labeling method described in \S~\ref{ref:finetunedataset} to assign fine-grained labels (e.g., use-before-initialization, null pointer dereference) beyond the original OOB and UAF labels to the same train set of 10,540 patches.
When expanding to a broader set of vulnerability types, including memory out-of-bounds access, use-after-free, null pointer dereference, use-before-initialization, memory leak and non-memory-related bugs, \system maintains robust performance with an \textbf{80.35\%} accuracy and an F1-score of 0.801. This demonstrates our method's ability to handle \textbf{diverse vulnerability types} while maintaining high classification accuracy.

In addition, even though we focused on the Linux kernel due to its vast scale and complexity, we also tested two other open-source projects: FFmpeg~\cite{ffmpeg} and OpenSSL~\cite{openssl}. 
Specifically, we followed the same data collection and labeling methodology described in \S~\ref{ref:finetunedataset}: we collected CVE patches for these projects by retrieving entries from the corresponding CVE database, automatically mapping their associated CWE labels into our bug type taxonomy, and then applied the same verification process to correct any mislabeled or ambiguous cases.
The results are equally compelling, with \system achieving F1-scores of 0.832 and 0.834 respectively, demonstrating its \textbf{generalizability across software}.

\subsubsection{Ablation Study}
\label{subsubsec:ablation}

The effectiveness of \system stems from several key design strategies. 
Through an ablation study, we now demonstrate how these design choices contribute to the overall performance. 

\begin{table}[]
\centering
{
\setlength{\tabcolsep}{3.5pt}
\begin{tabular}{|l|l|l|l|l|l|}
\hline
Method       & Accuracy & Precision & F1-score & FP rate & FN rate \\ \hline
LLM-only     & 76.7\%   & 81.1\%     & 0.776    & 12.6\%   & 19.0\%   \\ \hline
SliceLM-only & 76.0\%    & 80.0\%     & 0.769    & 10.1\%   & 16.4\%   \\ \hline
\system       & 87.4\%    & 87.7\%     & 0.875    & 5.4\%    & 10.0\%   \\ \hline
\end{tabular}
}
\caption{Comparison between LLM-only, SliceLM-only and \system}
\label{table:endtoend}
\vspace{-.2in}
\end{table}

\begin{table}[]
\centering
{
\setlength{\tabcolsep}{3pt}
\begin{tabular}{|l|l|l|l|l|l|}
\hline
Method                   & Accuracy & Precision & F1-score & FP rate & FN rate \\ \hline
3-line contexts                                                                   & 54.5\%   & 60.1\%    & 0.531    & 21.5\%  & 39.3\%  \\ \hline
Standard slices                                                                   & 58.1\%   & 72.1\%    & 0.635    & 29.1\%  & 56.0\%  \\ \hline
Selective slices & 67.0\%   & 70.5\%    & 0.680   & 16.5\%  & 34.1\%  \\ \hline
\begin{tabular}[c]{@{}l@{}}Selective slices+\\function renaming\end{tabular} & 78.0\%   & 81.5\%    & 0.789   & 10.5\%  & 22.5\%  \\ \hline
SliceLM slices      & 81.8\%   & 83.5\%    & 0.821    & 8.8\%   & 18.8\%  \\ \hline
\end{tabular}
}
\caption{The performance of \tool with different context options}
\vspace{-.3in}
\label{table:contextchoices}
\end{table}

\PP{Dual-method design} Our first experiment 
targets understanding whether the LLM-based method or \tool alone can achieve good results by themselves (similar to that of \system).
To test the LLM-only solution, we adapt the original LLM-based classification to always force an answer, instead of allowing the LLM to defer the answer when it deems that there are insufficient indicators in the patch. To test the \tool-only solution, we simply feed all the patches to \tool without any modifications to the model.
\autoref{table:endtoend} shows the results.
Interestingly, both the LLM-only and \tool-only solutions yield promising results, with similar performance across metrics such as accuracy, precision, etc.
\textit{Notably both already outperform SID and TreeVul, but are inferior to \system.}

Combining the two methods yield improved results across the board, e.g., an \textbf{11\%} improvement in overall accuracy. 
These results demonstrate the complementary nature of the two methods.
Particularly noteworthy is \system's improvement on false positives and false negatives simultaneously. It reduces the false positive rate to 5.4\% (compared to 12.6\% for LLM-only and 10.1\% for SliceLM-only) and the false negative rate to 10.0\% (versus 19.0\% and 16.4\% respectively).

\PP{Context options.} Code context is particularly crucial for the classification of use-after-free, memory out-of-bounds access and other memory corruption patches since their differences are subtle, as discussed previously. Thus, our ablation study on context options is done for this challenging fine-grained classification task (the second stage of our two-stage classification).
~\autoref{table:contextchoices} depicts the results.
The three lines before and after patch diffs as code context (the conventional and default way~\cite{context3lines}) proves inadequate, achieving only \textbf{54.5\%} accuracy.
Traditional program dependence based slices show a moderate improvement to yield an accuracy of \textbf{58.1\%}, but still fall short of capturing crucial vulnerability patterns.

Our custom slicing consists of three components: selective slicing, function renaming, and pruning of redundant code changes. Here, we seek to understand their individual and cumulative contributions. Selective slicing alone significantly improves performance over traditional program slicing, achieving 67.0\% accuracy compared to 58.1\%, thanks to improved dependency selection and noise reduction. Adding function renaming to selective slicing further improves performance to 78.0\% accuracy and an F1-score of 0.789. This confirms that normalizing semantically equivalent memory-related functions improves the model’s ability to generalize across diverse patches. Finally, our full slicing pipeline—which includes selective slicing, function renaming, and pruning of redundant code changes—achieves 81.8\% accuracy with an F1-score of 0.821. Notably, the complete pipeline reduces the false positive rate from 16.5\% (with selective slicing) to 8.8\%, and the false negative rate from 34.1\% to 18.8\%. These results demonstrate that each component meaningfully enhances the effectiveness of code context analysis and collectively enables the model to better distinguish subtle variations in memory corruption patches. Note that these results correspond to the performance of \tool and not the entire \system pipeline.

\PP{Pass better context to the LLM}
Now that we have demonstrated the importance of the right code context, a natural question arises: Can the LLM achieve performance similar to \system if given access to our custom slices? 
To test this,
we provide the LLM with carefully crafted prompts and examples demonstrating how to leverage the slice information.
Interestingly, it achieves only a \textbf{62.6\%} accuracy, with an F1-score of 0.664. This stands in stark contrast to \tool's superior performance of 81.8\% accuracy and 0.821 F1-score. We conjecture that this is due to the LLM not being familiar with our custom slice format, which does not always conform to well-formed programs. This is in contrast to \tool, which is pre-trained and fine-tuned on our specialized slices.

To circumvent this issue, we perform another experiment wherein we feed the entire function as code context to the LLM. In theory, the LLM should automatically extract the needed information from the whole function.
The results were better, but with only a \textbf{73.1\%} accuracy and an F1-score of 0.732.
Thus, there is still a gap compared to \tool's 81.8\% accuracy and 0.821 F1-score.
We suspect that this is because the markedly larger code context (whole functions) introduces noises and disrupts the LLM's understanding of a patch.

\PP{Two-stage classification for \difficult}\label{sec:two-stage}
To validate the effectiveness of the two-stage classification of patches without hints, we compare the approach with a single-stage classification where we directly classify patches into OOB, UAF, and others, without first classifying them into memory corruption vs. non-memory ones. While direct one-stage classification achieves a \textbf{71.4\%} accuracy with a 77.2\% precision and a 0.738 F1-score, our two-stage approach significantly improves performance to \textbf{77.9\%} accuracy, 78.4\% precision, and a 0.776 F1-score. These results demonstrate that decomposing the classification task into smaller sub-problems improves accuracy.

\PP{Using other LLMs}
To demonstrate that our approach is not dependent on a specific LLM, we also evaluated \system by using the popular open source  Llama 3.1 405B. Overall, on the same testing dataset, we find that the overall pipeline still maintains strong performance, achieving \textbf{85.0\%} accuracy, 85.0\% precision with an F1-score of 0.849.
It shows only a modest decrease from \system with GPT-4-turbo. 
A detailed analysis reveals interesting differences between the models. 
Notably, when using identical prompts, Llama classifies fewer patches into the category with hints--65\% versus 77\% with GPT-4-turbo.
Yet, its accuracy on such patches is only 88.0\% compared to 90.3\% in GPT-4-turbo. 
This indicates that Llama is more conservative in identifying patches with clear indicators.
Interestingly, Llama defers more patches without hints to \tool, but \tool can still achieve 79.4\% accuracy in these cases, which is higher than the result when GPT-4-turbo is employed (77.9\%).  
We also evaluated \system using the OpenAI-O1 model still using identical prompts. However, due to its high API cost, we only test 100 cases. In these cases, the overall performance is a \textbf{89.0\%} accuracy, a 89.3\% precision with a 0.890 F1-score. 
As with GPT-4-turbo, 23\% of the cases are classified as patches without hints, and \tool achieves a 87\% accuracy. 77\% of the cases are classified as patches with hints; OpenAI-O1 has an 89.6\% accuracy with respect to these.
Overall, these results show that while the model choice affects specific performance metrics, our approach remains effective across other LLMs, validating its robustness  and generality.

\subsubsection{Robustness of Function Renaming}A potential concern is that function renaming based on key memory semantics may suffer from 'semantic drift' across different kernel versions; for example, a function may increase a reference counter in one version but not in another. To assess this, we manually analyzed 100 randomly selected functions from the Linux kernel, comparing their key memory semantics in two versions spanning five years: v4.19 (October 2018) and v6.6 (October 2023). We found that only 3\% of the functions exhibited substantial semantic changes affecting memory-related behavior (e.g., transitioning from not freeing memory to including a \texttt{kfree()} operation). In contrast, 37\% showed no memory-related changes, 26\% involved only minor modifications (such as temporary variable refactoring) that do not affect semantic labeling, and 34\% were absent in one version due to removal or major API changes. These findings suggest that semantic drift poses minimal risk to our renaming strategy in practice.

\subsubsection{Analysis of the misclassified cases}
It is inherently challenging to precisely interpret the failures of machine learning models. However, we make some observations on the cases in which \system fails, in an attempt to contemplate how our solution can be further improved.
For patches with hints (analyzed by the LLM component), nearly all misclassifications arise from inaccuracies in determining whether patches contain reliable hints. This can be attributed to hallucination phenomena in LLMs, where the model either incorrectly perceives or overlooks critical indicators. Specifically, two distinct error patterns emerge: (1) patches lack clear signals about the bug type and necessitate deeper code context analysis, yet the LLM mistakenly perceives that sufficient hints are present, leading to incorrect inference; (2) patches explicitly or implicitly contain hints about the bug type, but the LLM fails to accurately extract or interpret these signals. An illustrative case is provided in Appendix~\S\ref{sec:duallmfail}.
For patches without hints (analyzed by the SliceLM component), more than 50\% of the misclassified cases exhibit complex structures in the patches. Some patches involve extensive modifications, such as removing a specific feature with hundreds of lines of code (sometimes spanning multiple files), and the vulnerable code is mixed in and removed at the same time. We provide a representative example in Appendix~\S\ref{sec:duallmfail}. In other cases, the key logic that helps us understand the vulnerability type and the actual patch location are separated by multiple function calls in the program's call graph, requiring a far larger analysis scope than what we can capture (even with the help of function renaming). Interestingly, about 10\% of the misclassified patches occur in cases where, from a human analyst's perspective, the patch diffs and slices contain sufficient information to determine the vulnerability type; yet, the model fails to make the correct classification. These cases highlight the gap between human and machine comprehension of patch features. The remaining failures stem from various other factors that we are unable to fully identify or systematically categorize.

\subsubsection{Data leakage concerns with LLM}
To address potential concerns about LLM performance being influenced by exposure to CVE patches during pre-training, we compared the results on patches before and after the LLM's training cutoff date (2024). On pre-2024 patches, \system achieves 87.9\% accuracy, and a 0.880 F1-score. For patches from 2024, it achieves 82.1\% accuracy, and a 0.820 F1-score. 
While there is a moderate drop in performance,  \system still maintains reasonably strong performance, suggesting its effectiveness stems from its fundamental design rather than memorization of training examples. The system can generalize to unseen patches while maintaining robust performance.

\subsection{Evaluation on Unlabeled Recent Kernel Patches}
\label{sec:randomeval}

\subsubsection{Comparative study}
Besides the CVE patches, we also evaluated 5,140 recent patches collected from the Linux kernel 6.6 LTS branch in 2024 (details in \S\ref{sec:setup}). These patches include both security and non-security ones.
Because \system requires security patches as input, we evaluated it in combination with different upstream binary classifiers. Specifically, we used GraphSPD~\cite{wang2022graphspd} and VulFixMiner~\cite{zhou2021finding} to identify security patches from 5,140 recent patches. 
As for PatchRNN, we found it unsuitable for this task. It labeled nearly 97\% of test cases as “security.” We manually verified 50 cases and found that from these predicted “security” related patches only 22\% were true positives, demonstrating low precision and thus, limiting its practicality for downstream fine-grained classification.

GraphSPD identified 403 security patches, of which \system classified 142 as UAF/OOB. Manual verification confirmed that 90 of these are true UAF/OOB patches, yielding an overall precision of \textbf{63.4\%}. If 31 false positives were excluded due to incorrect GraphSPD classifications (that is, these were not actually security-related as verified manually), we would retain 111 candidates, resulting in a filtered precision (excluding GraphSPD errors) of \textbf{ 81.1\%}. Using VulFixMiner as the binary classifier, 110 security patches were identified, and DUALLM reported 39 UAF/OOB cases. Among them, 20 were true positives and 16 were non-security patches, resulting in an overall precision of 51.3\%, and a filtered precision of 86.9\%.

We also evaluated TreeVul with the same settings for comparison. GraphSPD + TreeVul yielded 104 UAF/OOB predictions, with only 25 true positives and 51 non-security cases, resulting in 24.0\% overall precision and 47.2\% filtered precision. With VulFixMiner + TreeVul, 51 patches were classified as UAF/OOB, of which 20 were correct and 18 were non-security, yielding a 39.2\% overall precision and a 60.6\% filtered precision.

Lastly, we include SID for comparison, which does not require a separate binary classifier. SID correctly identified 42 UAF / OOB patches with a precision of 78\%, but missed the majority due to its limited coverage of the rules.

\begin{figure}[h]
\centering
  \includegraphics[width=0.3\textwidth]{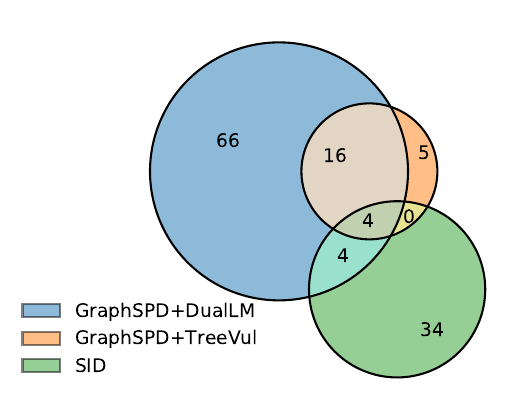}
  \caption{The Venn diagram for GraphSPD+DualLM, GraphSPD+TreeVul and SID}
  \label{fig:venn1}
\vspace{-.2in}
\end{figure}

\begin{figure}[h]
\centering
  \includegraphics[width=0.3\textwidth]{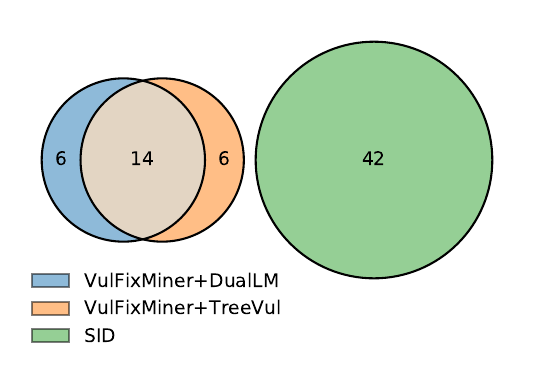}
  \caption{The Venn diagram for VulFixMiner+DualLM, VulFixMiner+TreeVul and SID}
  \label{fig:venn2}
\vspace{-.3in}
\end{figure}

\subsubsection{True positive study}
To better understand the overlap between true positives identified by different pipelines, we constructed Venn diagrams (\autoref{fig:venn1} and ~\autoref{fig:venn2}) based on combinations of binary classifiers (GraphSPD or VulFixMiner) and fine-grained vulnerability classifiers (\system or TreeVul), with SID included for reference. For ease of visual exposition, we split the comparisons by binary classifier. In ~\autoref{fig:venn1}, we show results for GraphSPD-based pipelines and SID. \system identifies 90 true UAF/OOB positives, TreeVul identifies 25, and SID identifies 42. Notably, 64\% (20 out of 25) of TreeVul’s true positives overlap with \system, showing that most of the true positives identified by TreeVul can also be identified by \system. Manual inspection of TreeVul-only positives reveals that these cases can be reliably classified using only the patch diffs—TreeVul's sole input—while additional contextual information (e.g., slicing) may introduce noise. The limited overlap with SID stems largely from the limited agreement between SID and GraphSPD (only 8 overlapping patches are identified by GraphSPD as security).

~\autoref{fig:venn2} presents a similar comparison using VulFixMiner as the binary classifier. Here, \system and TreeVul identify 20 true positives each, with 14 in common, and in this case, SID  contributes a fully disjoint set of 42 true positives. The absence of any overlap between SID and VulFixMiner explains the disconnect in the results.
Specifically, the disparity arises because VulFixMiner targets Java and Python projects, while SID—like our work—focuses on C/C++ codebases such as the Linux kernel.
For the same reason, i.e., VulFixMiner's difficulty with C/C++ codebases, the subsequent results with both \system and TreeVul are more modest compared to that with GraphSPD. In other words, as one might expect, the quality of binary classifiers can significantly affect follow-up results with a fine-grained classifier.

\subsubsection{False negative study}
While the Venn diagrams highlight the coverage and overlap of different pipelines in identifying true positives, it is equally important to understand the limitations of our system—particularly its false negatives. To this end, we further investigate the negative cases from the GraphSPD + DUALLM pipeline to assess whether any true UAF or OOB patches were missed. 
We randomly sample 100 patches (representing approximately 40\% of the total patches labeled as non-UAF-OOB) labeled as non-UAF-OOB and find one false negative (possibly due to a hallucination of the LLM), yielding a \textbf{1\%} false negative rate.

\subsubsection{Case study}
Interestingly, while 8 of the 90 UAF/OOB patches identified by GraphSPD + \system pipeline, fix bugs that were reported by Syzbot~\cite{syzbot} (a continuous fuzzing platform that has reported thousands of Linux kernel bugs), these bugs were not reported as use-after-free or memory out-of-bounds vulnerabilities.
Developing reproducers or exploits for kernel vulnerabilities is challenging, particularly since many cases involve race conditions or hardware-specific contexts (such as Dell or NVIDIA devices). In spite of this, 
we picked two of the syzbot bugs that are not reported as UAF or OOB, and successfully modified the reproducers reported on syzbot to get the PoCs confirming the UAF and OOB behaviors.
Most notably, we also developed one exploit against one UAF bug (out of the two cases) which can successfully achieve a \textbf{control flow hijacking attack}. This demonstrates the real-world impact of our approach.

\begin{figure}[t]
    \centering
    \begin{minipage}[t]{1\linewidth}
        \begin{lstlisting}[]

The patch fdb8e12cc2cc ("netfilter: ipset: fix performance regression in swap operation") missed to add the calls to gc cancellations at the error path of create operations and at module unload. Also, because the half of the destroy operations now executed by a function registered by call_rcu(), neither NFNL_SUBSYS_IPSET mutex or rcu read lock is held and therefore the checking of them results false warnings.

 ...
@@ -2378,6 +2379,7 @@ ip_set_net_exit(struct net *net)
 		set = ip_set(inst, i);
 		if (set) {
 			ip_set(inst, i) = NULL;
+			set->variant->cancel_gc(set);
 			ip_set_destroy_set(set);
 		}
 	}
  ...
\end{lstlisting}
    \end{minipage}
    \caption{A use-after-free patch identified by \system}
    \vspace{-.2in}
    \label{fig:patch27c5a09}
\end{figure}

\begin{figure}[t]
    \centering
    \begin{minipage}[t]{1\linewidth}
        \begin{lstlisting}[]
static int ip_set_create(struct sk_buff *skb, const struct nfnl_info *info, ...) {
 ...
 cleanup:
+   set->variant->cancel_gc(set);
 	 set->variant->destroy(set);
 ...
}

// free `h`
static void  mtype_destroy(struct ip_set *set) {
    struct htype *h = set->data;
    ...
    kfree(h);
    set->data = NULL;
}

static void mtype_cancel_gc(struct ip_set *set) {
    struct htype *h = set->data;
    if (SET_WITH_TIMEOUT(set))
        cancel_delayed_work_sync(&h->gc.dwork);
}

static void expire_timers(struct timer_base *base, struct hlist_head *head) {
    ...
    struct timer_list *timer;
    void (*fn)(struct timer_list *);
    // timer is a dangling pointer
    timer = hlist_entry(head->first, struct timer_list, entry);  
    fn = timer->function;
    ...
    // Control flow hijacking
    call_timer_fn(timer, fn, baseclk); 
}
\end{lstlisting}
    \end{minipage}
    \vspace{-.2in}
    \caption{Exploitable primitive}
    \vspace{-.2in}
    \label{fig:capability}
\end{figure}

Next, we describe how we develop the control flow hijacking attack to exploit the UAF bug.
Figure~\ref{fig:patch27c5a09} illustrates a patch addressing a bug identified by \system as a use-after-free vulnerability. Note that the commit title/message and code diffs do not directly reveal the bug type, and therefore it was ultimately classified by \tool.

As shown in Figure~\ref{fig:capability}, the function call to \texttt{mtype_destroy()} on line 5 ultimately frees the \texttt{htype} object on line 13. Without \texttt{cancel_gc()} being invoked on line 4 —which calls \texttt{mtype_cancel_gc()} to terminate the worker process —the worker process would eventually be triggered after a timeout. Because the \texttt{htype} object has already been freed, an attacker could exploit this UAF vulnerability by performing heap spraying to gain control of the freed \texttt{htype} memory. Specifically, the \texttt{timer} pointer, derived on line 28, points to the freed memory. By overwriting \texttt{timer->function}, which is a function pointer, an attacker can achieve a control flow hijack on line 32 when it is de-referenced.
We developed a working exploit that sprayed \texttt{user_key_payload} objects (which are elastic) in kmalloc-2k slabs, to achieve the overwrite and successfully achieved the control flow hijack.

\section{Conclusion}

In this paper, we presented \system, a Dual-method approach for identifying patches fixing critical memory bugs(e.g., memory out-of-bounds access and use-after-free). Based on our observation that patches can be naturally divided into those with and without clear indicators, \system strategically combines an LLM's capabilities to understand natural language and recognize patch patterns, with a specialized model (\tool) trained on our custom slices.  
Our comprehensive evaluations demonstrate \system's effectiveness. It achieves an 87.4\% accuracy on a quality-controlled CVE dataset, significantly outperforming existing approaches. Most importantly, in analyzing 5,140 recent patches, \system identified 90 memory out-of-bounds and use-after-free patches, with two proof-of-concept programs and one exploit achieving successful control flow hijacking, confirming their severity. This underscores the practical impact of our system in identifying crucial patches.
\clearpage

\bibliography{reference}

@String { COMPUTER     = {IEEE Computer Magazine} }

@String { JUN          = {June} }

@article{li2024investigation,
  title={An Investigation of Patch Porting Practices of the Linux Kernel Ecosystem},
  author={Li, Xingyu and Zhang, Zheng and Qian, Zhiyun and Jaeger, Trent and Song, Chengyu},
  journal={arXiv preprint arXiv:2402.05212},
  year={2024}
}

@article{ahmad2021unified,
  title={Unified pre-training for program understanding and generation},
  author={Ahmad, Wasi Uddin and Chakraborty, Saikat and Ray, Baishakhi and Chang, Kai-Wei},
  journal={arXiv preprint arXiv:2103.06333},
  year={2021}
}

@misc{li2023hitchhikers,
      title={The Hitchhiker's Guide to Program Analysis: A Journey with Large Language Models}, 
      author={Haonan Li and Yu Hao and Yizhuo Zhai and Zhiyun Qian},
      year={2023},
      eprint={2308.00245},
      archivePrefix={arXiv},
      primaryClass={cs.SE}
}

@inproceedings{pan2023fine,
  title={Fine-grained Commit-level Vulnerability Type Prediction by CWE Tree Structure},
  author={Pan, Shengyi and Bao, Lingfeng and Xia, Xin and Lo, David and Li, Shanping},
  booktitle={2023 IEEE/ACM 45th International Conference on Software Engineering (ICSE)},
  pages={957--969},
  year={2023},
  organization={IEEE}
}

@inproceedings{zhang2021investigation,
  title={An Investigation of the Android Kernel Patch Ecosystem},
  author={Zhang, Zheng and Zhang, Hang and Qian, Zhiyun and Lau, Billy},
  booktitle={30th $\{$USENIX$\}$ Security Symposium ($\{$USENIX$\}$ Security 21)},
  year={2021}
}

@Misc{submitpatchrules,
  Title                    = {{Submitting patches: the essential guide to getting your code into the kernel}},
  HowPublished             = {\url{https://www.kernel.org/doc/html/next/process/submitting-patches.html}},
}

@book{bishop2006pattern,
  title={Pattern recognition and machine learning},
  author={Bishop, Christopher M and Nasrabadi, Nasser M},
  volume={4},
  number={4},
  year={2006},
  publisher={Springer}
}

@Misc{cve-intro,
  Title                    = {{CVEs and the NVD Process}},
  HowPublished             = {\url{https://nvd.nist.gov/general/cve-process#:~:text=Founded%20in%201999%2C%20the%20CVE,Infrastructure%20Security%20Agency%20(CISA).}},
}

@Misc{linuxcoding,
  Title                    = {{Linux Kernel Coding Style}},
  author={Linus, Torvalds},
  HowPublished             = {\url{https://slurm.schedmd.com/coding_style.pdf}},
}

@article{zhao2023survey,
  title={A survey of large language models},
  author={Zhao, Wayne Xin and Zhou, Kun and Li, Junyi and Tang, Tianyi and Wang, Xiaolei and Hou, Yupeng and Min, Yingqian and Zhang, Beichen and Zhang, Junjie and Dong, Zican and others},
  journal={arXiv preprint arXiv:2303.18223},
  year={2023}
}

@inproceedings{wang2022graphspd,
  title={GraphSPD: Graph-Based Security Patch Detection with Enriched Code Semantics},
  author={Wang, Shu and Wang, Xinda and Sun, Kun and Jajodia, Sushil and Wang, Haining and Li, Qi},
  booktitle={2023 IEEE Symposium on Security and Privacy (SP)},
  pages={604--621},
  year={2022},
  organization={IEEE Computer Society}
}

@inproceedings{wu2020precisely,
  title={Precisely Characterizing Security Impact in a Flood of Patches via Symbolic Rule Comparison.},
  author={Wu, Qiushi and He, Yang and McCamant, Stephen and Lu, Kangjie},
  booktitle={NDSS},
  year={2020}
}

@inproceedings{li2017large,
  title={A large-scale empirical study of security patches},
  author={Li, Frank and Paxson, Vern},
  booktitle={Proceedings of the 2017 ACM SIGSAC Conference on Computer and Communications Security},
  pages={2201--2215},
  year={2017}
}

@inproceedings{dunlap2024vfcfinder,
  title={VFCFinder: Pairing Security Advisories and Patches},
  author={Dunlap, Trevor and Lin, Elizabeth and Enck, William and Reaves, Bradley},
  booktitle={Proceedings of the 19th ACM Asia Conference on Computer and Communications Security},
  pages={1128--1142},
  year={2024}
}

@article{vaswani2017attention,
  title={Attention is all you need},
  author={Vaswani, Ashish and Shazeer, Noam and Parmar, Niki and Uszkoreit, Jakob and Jones, Llion and Gomez, Aidan N and Kaiser, {\L}ukasz and Polosukhin, Illia},
  journal={Advances in neural information processing systems},
  volume={30},
  year={2017}
}

@Misc{ffmpeg,
  Title                    = {{FFmpeg}},
  HowPublished             = {\url{https://ffmpeg.org/}},
}

@Misc{openssl,
  Title                    = {{OpenSSL}},
  HowPublished             = {\url{https://www.openssl.org/}},
}

@inproceedings{wu2018fuze,
  title={$\{$FUZE$\}$: Towards facilitating exploit generation for kernel use-after-free vulnerabilities},
  author={Wu, Wei and Chen, Yueqi and Xu, Jun and Xing, Xinyu and Gong, Xiaorui and Zou, Wei},
  booktitle={27th $\{$USENIX$\}$ Security Symposium ($\{$USENIX$\}$ Security 18)},
  pages={781--797},
  year={2018}
}

@inproceedings{chen2020koobe,
  title={KOOBE: Towards facilitating exploit generation of kernel out-of-bounds write vulnerabilities},
  author={Chen, Weiteng and Zou, Xiaochen and Li, Guoren and Qian, Zhiyun},
  booktitle={Proceedings of the 29th USENIX Conference on Security Symposium},
  pages={1093--1110},
  year={2020}
}

@Misc{joerdoc,
  Title                    = {{Joern Documentation}},
  HowPublished             = {\url{https://docs.joern.io/home}},
}

@inproceedings{yamaguchi2014modeling,
  title={Modeling and discovering vulnerabilities with code property graphs},
  author={Yamaguchi, Fabian and Golde, Nico and Arp, Daniel and Rieck, Konrad},
  booktitle={2014 IEEE Symposium on Security and Privacy},
  pages={590--604},
  year={2014},
  organization={IEEE}
}

@Misc{openai2023gpt4,
      title={GPT-4 Technical Report}, 
      author={OpenAI},
      year={2023},
      eprint={2303.08774},
      archivePrefix={arXiv},
      primaryClass={cs.CL}
}

@Misc{openaio1,
  Title                    = {{OpenAI O1 system card}},
  HowPublished             = {\url{https://cdn.openai.com/o1-system-card.pdf/}},
}

@Misc{gpt4turbo,
  Title                    = {{GPT-4 Turbo in the OpenAI API}},
  HowPublished             = {\url{https://help.openai.com/en/articles/8555510-gpt-4-turbo-in-the-openai-api/}},
}

@Misc{Llamaintro,
  Title                    = {{Introducing LLaMA: A foundational, 65-billion-parameter large language model}},
  HowPublished             = {\url{https://ai.facebook.com/blog/large-language-model-llama-meta-ai/}},
}

@Misc{Llama2intro,
  Title                    = {{Llama 2}},
  HowPublished             = {\url{https://llama.meta.com/llama2}},
}

@Misc{Llama31intro,
  Title                    = {{Llama 3.1}},
  HowPublished             = {\url{https://ai.meta.com/blog/meta-llama-3-1/}},
}

@misc{feng2023prompting,
      title={Prompting Is All Your Need: Automated Android Bug Replay with Large Language Models}, 
      author={Sidong Feng and Chunyang Chen},
      year={2023},
      eprint={2306.01987},
      archivePrefix={arXiv},
      primaryClass={cs.SE}
}

@Misc{chagptintro,
  Title                    = {{Introducing ChatGPT}},
  HowPublished             = {\url{https://openai.com/blog/chatgpt}},
}

@misc{wang2023defecthunternovelllmdrivenboostedconformerbased,
      title={DefectHunter: A Novel LLM-Driven Boosted-Conformer-based Code Vulnerability Detection Mechanism}, 
      author={Jin Wang and Zishan Huang and Hengli Liu and Nianyi Yang and Yinhao Xiao},
      year={2023},
      eprint={2309.15324},
      archivePrefix={arXiv},
      primaryClass={cs.CR},
      url={https://arxiv.org/abs/2309.15324}, 
}

@misc{alrashedy2024llmspatchsecurityissues,
      title={Can LLMs Patch Security Issues?}, 
      author={Kamel Alrashedy and Abdullah Aljasser and Pradyumna Tambwekar and Matthew Gombolay},
      year={2024},
      eprint={2312.00024},
      archivePrefix={arXiv},
      primaryClass={cs.CR},
      url={https://arxiv.org/abs/2312.00024}, 
}

@misc{tang2024justintimedetectionsilentsecurity,
      title={Just-in-Time Detection of Silent Security Patches}, 
      author={Xunzhu Tang and Zhenghan Chen and Kisub Kim and Haoye Tian and Saad Ezzini and Jacques Klein},
      year={2024},
      eprint={2312.01241},
      archivePrefix={arXiv},
      primaryClass={cs.CR},
      url={https://arxiv.org/abs/2312.01241}, 
}

@misc{sheng2024lprotectorllmdrivenvulnerabilitydetection,
      title={LProtector: An LLM-driven Vulnerability Detection System}, 
      author={Ze Sheng and Fenghua Wu and Xiangwu Zuo and Chao Li and Yuxin Qiao and Lei Hang},
      year={2024},
      eprint={2411.06493},
      archivePrefix={arXiv},
      primaryClass={cs.CR},
      url={https://arxiv.org/abs/2411.06493}, 
}

@misc{ullah2024llmsreliablyidentifyreason,
      title={LLMs Cannot Reliably Identify and Reason About Security Vulnerabilities (Yet?): A Comprehensive Evaluation, Framework, and Benchmarks}, 
      author={Saad Ullah and Mingji Han and Saurabh Pujar and Hammond Pearce and Ayse Coskun and Gianluca Stringhini},
      year={2024},
      eprint={2312.12575},
      archivePrefix={arXiv},
      primaryClass={cs.CR},
      url={https://arxiv.org/abs/2312.12575}, 
}

@inproceedings{ahmed2023better,
  title={Better patching using LLM prompting, via Self-Consistency},
  author={Ahmed, Toufique and Devanbu, Premkumar},
  booktitle={2023 38th IEEE/ACM International Conference on Automated Software Engineering (ASE)},
  pages={1742--1746},
  year={2023},
  organization={IEEE}
}

@article{yang2310white,
  title={White-box compiler fuzzing empowered by large language models. CoRR, abs/2310.15991, 2023b. doi: 10.48550},
  author={Yang, Chenyuan and Deng, Yinlin and Lu, Runyu and Yao, Jiayi and Liu, Jiawei and Jabbarvand, Reyhaneh and Zhang, Lingming},
  journal={arXiv preprint ARXIV.2310.15991}
}

@misc{yang2024kernelgptenhancedkernelfuzzing,
      title={KernelGPT: Enhanced Kernel Fuzzing via Large Language Models}, 
      author={Chenyuan Yang and Zijie Zhao and Lingming Zhang},
      year={2024},
      eprint={2401.00563},
      archivePrefix={arXiv},
      primaryClass={cs.CR},
      url={https://arxiv.org/abs/2401.00563}, 
}

@article{radford2019language,
  title={Language models are unsupervised multitask learners},
  author={Radford, Alec and Wu, Jeffrey and Child, Rewon and Luan, David and Amodei, Dario and Sutskever, Ilya and others},
  journal={OpenAI blog},
  volume={1},
  number={8},
  pages={9},
  year={2019}
}

@inproceedings{wang2021patchdb,
  title={Patchdb: A large-scale security patch dataset},
  author={Wang, Xinda and Wang, Shu and Feng, Pengbin and Sun, Kun and Jajodia, Sushil},
  booktitle={2021 51st Annual IEEE/IFIP International Conference on Dependable Systems and Networks (DSN)},
  pages={149--160},
  year={2021},
  organization={IEEE}
}

@article{wei2022chain,
  title={Chain of thought prompting elicits reasoning in large language models},
  author={Wei, Jason and Wang, Xuezhi and Schuurmans, Dale and Bosma, Maarten and Chi, Ed and Le, Quoc and Zhou, Denny},
  journal={arXiv preprint arXiv:2201.11903},
  year={2022}
}

@article{qin2023chatgpt,
  title={Is ChatGPT a general-purpose natural language processing task solver?},
  author={Qin, Chengwei and Zhang, Aston and Zhang, Zhuosheng and Chen, Jiaao and Yasunaga, Michihiro and Yang, Diyi},
  journal={arXiv preprint arXiv:2302.06476},
  year={2023}
}

@inproceedings{zhou2017automated,
  title={Automated identification of security issues from commit messages and bug reports},
  author={Zhou, Yaqin and Sharma, Asankhaya},
  booktitle={Proceedings of the 2017 11th joint meeting on foundations of software engineering},
  pages={914--919},
  year={2017}
}

@article{lewis2019bart,
  title={Bart: Denoising sequence-to-sequence pre-training for natural language generation, translation, and comprehension},
  author={Lewis, Mike and Liu, Yinhan and Goyal, Naman and Ghazvininejad, Marjan and Mohamed, Abdelrahman and Levy, Omer and Stoyanov, Ves and Zettlemoyer, Luke},
  journal={arXiv preprint arXiv:1910.13461},
  year={2019}
}

@article{devlin2018bert,
  title={Bert: Pre-training of deep bidirectional transformers for language understanding},
  author={Devlin, Jacob and Chang, Ming-Wei and Lee, Kenton and Toutanova, Kristina},
  journal={arXiv preprint arXiv:1810.04805},
  year={2018}
}

@inproceedings{yu2020order,
  title={Order matters: Semantic-aware neural networks for binary code similarity detection},
  author={Yu, Zeping and Cao, Rui and Tang, Qiyi and Nie, Sen and Huang, Junzhou and Wu, Shi},
  booktitle={Proceedings of the AAAI Conference on Artificial Intelligence},
  volume={34},
  number={01},
  pages={1145--1152},
  year={2020}
}

@misc{sawadogo2020learningcatchsecuritypatches,
      title={Learning to Catch Security Patches}, 
      author={Arthur D. Sawadogo and Tegawendé F. Bissyandé and Naouel Moha and Kevin Allix and Jacques Klein and Li Li and Yves Le Traon},
      year={2020},
      eprint={2001.09148},
      archivePrefix={arXiv},
      primaryClass={cs.SE},
      url={https://arxiv.org/abs/2001.09148}, 
}

@inproceedings{zhou2023colefunda,
  title={Colefunda: Explainable silent vulnerability fix identification},
  author={Zhou, Jiayuan and Pacheco, Michael and Chen, Jinfu and Hu, Xing and Xia, Xin and Lo, David and Hassan, Ahmed E},
  booktitle={2023 IEEE/ACM 45th International Conference on Software Engineering (ICSE)},
  pages={2565--2577},
  year={2023},
  organization={IEEE}
}

@article{fu2023vulexplainer,
  title={Vulexplainer: A transformer-based hierarchical distillation for explaining vulnerability types},
  author={Fu, Michael and Nguyen, Van and Tantithamthavorn, Chakkrit Kla and Le, Trung and Phung, Dinh},
  journal={IEEE Transactions on Software Engineering},
  year={2023},
  publisher={IEEE}
}

@misc{nguyen2023multigranularitydetectorvulnerabilityfixes,
      title={Multi-Granularity Detector for Vulnerability Fixes}, 
      author={Truong Giang Nguyen and Thanh Le-Cong and Hong Jin Kang and Ratnadira Widyasari and Chengran Yang and Zhipeng Zhao and Bowen Xu and Jiayuan Zhou and Xin Xia and Ahmed E. Hassan and Xuan-Bach D. Le and David Lo},
      year={2023},
      eprint={2305.13884},
      archivePrefix={arXiv},
      primaryClass={cs.CR},
      url={https://arxiv.org/abs/2305.13884}, 
}

@inproceedings{ganz2023pavudi,
  title={PAVUDI: Patch-based Vulnerability Discovery using Machine Learning},
  author={Ganz, Tom and Imgrund, Erik and H{\"a}rterich, Martin and Rieck, Konrad},
  booktitle={Proceedings of the 39th Annual Computer Security Applications Conference},
  pages={704--717},
  year={2023}
}

@inproceedings{zhou2021finding,
  title={Finding a needle in a haystack: Automated mining of silent vulnerability fixes},
  author={Zhou, Jiayuan and Pacheco, Michael and Wan, Zhiyuan and Xia, Xin and Lo, David and Wang, Yuan and Hassan, Ahmed E},
  booktitle={2021 36th IEEE/ACM International Conference on Automated Software Engineering (ASE)},
  pages={705--716},
  year={2021},
  organization={IEEE}
}

@misc{he2023bingoidentifyingsecuritypatches,
      title={BinGo: Identifying Security Patches in Binary Code with Graph Representation Learning}, 
      author={Xu He and Shu Wang and Pengbin Feng and Xinda Wang and Shiyu Sun and Qi Li and Kun Sun},
      year={2023},
      eprint={2312.07921},
      archivePrefix={arXiv},
      primaryClass={cs.CR},
      url={https://arxiv.org/abs/2312.07921}, 
}

@inproceedings{wang2021patchrnn,
  title={Patchrnn: A deep learning-based system for security patch identification},
  author={Wang, Xinda and Wang, Shu and Feng, Pengbin and Sun, Kun and Jajodia, Sushil and Benchaaboun, Sanae and Geck, Frank},
  booktitle={MILCOM 2021-2021 IEEE Military Communications Conference (MILCOM)},
  pages={595--600},
  year={2021},
  organization={IEEE}
}

@inproceedings{wang2019detecting,
  title={Detecting" 0-day" vulnerability: An empirical study of secret security patch in OSS},
  author={Wang, Xinda and Sun, Kun and Batcheller, Archer and Jajodia, Sushil},
  booktitle={2019 49th Annual IEEE/IFIP International Conference on Dependable Systems and Networks (DSN)},
  pages={485--492},
  year={2019},
  organization={IEEE}
}

@article{chen2021evaluating,
  title={Evaluating large language models trained on code},
  author={Chen, Mark and Tworek, Jerry and Jun, Heewoo and Yuan, Qiming and Pinto, Henrique Ponde de Oliveira and Kaplan, Jared and Edwards, Harri and Burda, Yuri and Joseph, Nicholas and Brockman, Greg and others},
  journal={arXiv preprint arXiv:2107.03374},
  year={2021}
}

@inproceedings{lemieux2023codamosa,
  title={CODAMOSA: Escaping coverage plateaus in test generation with pre-trained large language models},
  author={Lemieux, Caroline and Inala, Jeevana Priya and Lahiri, Shuvendu K and Sen, Siddhartha},
  booktitle={International conference on software engineering (ICSE)},
  year={2023}
}

@article{weiser1984program,
  title={Program slicing},
  author={Weiser, Mark},
  journal={IEEE Transactions on software engineering},
  number={4},
  pages={352--357},
  year={1984},
  publisher={IEEE}
}

@article{xia2023keep,
  title={Keep the Conversation Going: Fixing 162 out of 337 bugs for \$0.42 each using ChatGPT},
  author={Xia, Chunqiu Steven and Zhang, Lingming},
  journal={arXiv preprint arXiv:2304.00385},
  year={2023}
}

@inproceedings{pearce2022examining,
  title={Examining Zero-Shot Vulnerability Repair with Large Language Models},
  author={Pearce, Hammond and Tan, Benjamin and Ahmad, Baleegh and Karri, Ramesh and Dolan-Gavitt, Brendan},
  booktitle={2023 IEEE Symposium on Security and Privacy (SP)},
  pages={1--18},
  year={2022},
  organization={IEEE Computer Society}
}

@article{brown2020language,
  title={Language models are few-shot learners},
  author={Brown, Tom and Mann, Benjamin and Ryder, Nick and Subbiah, Melanie and Kaplan, Jared D and Dhariwal, Prafulla and Neelakantan, Arvind and Shyam, Pranav and Sastry, Girish and Askell, Amanda and others},
  journal={Advances in neural information processing systems},
  volume={33},
  pages={1877--1901},
  year={2020}
}

@article{radford2018improving,
  title={Improving language understanding by generative pre-training},
  author={Radford, Alec and Narasimhan, Karthik and Salimans, Tim and Sutskever, Ilya and others},
  year={2018},
  publisher={OpenAI}
}

@inproceedings{hoang2019patchnet,
  title={PatchNet: a tool for deep patch classification},
  author={Hoang, Thong and Lawall, Julia and Oentaryo, Richard J and Tian, Yuan and Lo, David},
  booktitle={2019 IEEE/ACM 41st International Conference on Software Engineering: Companion Proceedings (ICSE-Companion)},
  pages={83--86},
  year={2019},
  organization={IEEE}
}

@inproceedings{wen2019ptracer,
  title={PTracer: A Linux kernel patch trace bot},
  author={Wen, Yang and Cao, Jicheng and Cheng, Shengyu},
  booktitle={2019 34th IEEE/ACM International Conference on Automated Software Engineering (ASE)},
  pages={1210--1211},
  year={2019},
  organization={IEEE}
}

@article{feng2020codebert,
  title={Codebert: A pre-trained model for programming and natural languages},
  author={Feng, Zhangyin and Guo, Daya and Tang, Duyu and Duan, Nan and Feng, Xiaocheng and Gong, Ming and Shou, Linjun and Qin, Bing and Liu, Ting and Jiang, Daxin and others},
  journal={arXiv preprint arXiv:2002.08155},
  year={2020}
}

@article{guo2020graphcodebert,
  title={Graphcodebert: Pre-training code representations with data flow},
  author={Guo, Daya and Ren, Shuo and Lu, Shuai and Feng, Zhangyin and Tang, Duyu and Liu, Shujie and Zhou, Long and Duan, Nan and Svyatkovskiy, Alexey and Fu, Shengyu and others},
  journal={arXiv preprint arXiv:2009.08366},
  year={2020}
}

@Misc{context3lines,
  Title                    = {{Git -diff-config Documentation}},
  HowPublished             = {\url{https://git-scm.com/docs/diff-config}},
}

@Misc{wronglabelcve,
  Title                    = {{CVE-2016-5400}},
  HowPublished             = {\url{https://nvd.nist.gov/vuln/detail/cve-2016-5400}},
}

@Misc{wronglabelcwe,
  Title                    = {{CWE-119}},
  HowPublished             = {\url{https://cwe.mitre.org/data/definitions/119.html}},
}

@inproceedings{wang2018revery,
  title={Revery: From proof-of-concept to exploitable},
  author={Wang, Yan and Zhang, Chao and Xiang, Xiaobo and Zhao, Zixuan and Li, Wenjie and Gong, Xiaorui and Liu, Bingchang and Chen, Kaixiang and Zou, Wei},
  booktitle={Proceedings of the 2018 ACM SIGSAC conference on computer and communications security},
  pages={1914--1927},
  year={2018}
}

@article{xu2023autopwn,
  title={Autopwn: Artifact-assisted heap exploit generation for ctf pwn competitions},
  author={Xu, Dandan and Chen, Kai and Lin, Miaoqian and Lin, Chaoyang and Wang, Xiaofeng},
  journal={IEEE Transactions on Information Forensics and Security},
  year={2023},
  publisher={IEEE}
}

@inproceedings{syzscope,
  title={{SyzScope: Revealing High-Risk Security Impacts of Fuzzer-Exposed Bugs in Linux kernel}},
  author={Xiaochen Zou and Guoren Li and Weiteng Chen and Hang Zhang and Zhiyun Qian},
  booktitle={USENIX Security Symposium},
  year={2022}
}

@article{zhou2021spi,
  title={Spi: Automated identification of security patches via commits},
  author={Zhou, Yaqin and Siow, Jing Kai and Wang, Chenyu and Liu, Shangqing and Liu, Yang},
  journal={ACM Transactions on Software Engineering and Methodology (TOSEM)},
  volume={31},
  number={1},
  pages={1--27},
  year={2021},
  publisher={ACM New York, NY}
}

@inproceedings{zuo2023commit,
  title={Commit message can help: security patch detection in open source software via transformer},
  author={Zuo, Fei and Zhang, Xin and Song, Yuqi and Rhee, Junghwan and Fu, Jicheng},
  booktitle={2023 IEEE/ACIS 21st International Conference on Software Engineering Research, Management and Applications (SERA)},
  pages={345--351},
  year={2023},
  organization={IEEE}
}

@inproceedings{sun2023silent,
  title={Silent vulnerable dependency alert prediction with vulnerability key aspect explanation},
  author={Sun, Jiamou and Xing, Zhenchang and Lu, Qinghua and Xu, Xiwei and Zhu, Liming and Hoang, Thong and Zhao, Dehai},
  booktitle={2023 IEEE/ACM 45th International Conference on Software Engineering (ICSE)},
  pages={970--982},
  year={2023},
  organization={IEEE}
}

@Misc{CVEproblems,
  Title                    = {{What to do about CVE numbers}},
  Author                   = {Jonathan Corbet},
  HowPublished             = {\url{https://lwn.net/Articles/801157/}},
}

@Misc{syzbot,
  Title                    = {{Syzbot}},
  Author                   = {Google},
  HowPublished             = {\url{https://syzkaller.appspot.com/upstream/}},
}

@inproceedings{tian2012identifying,
  title={Identifying linux bug fixing patches},
  author={Tian, Yuan and Lawall, Julia and Lo, David},
  series = {ICSE'12},
}

@inproceedings{syzbridge,
  title={SyzBridge: Bridging the Gap in Exploitability Assessment of Linux Kernel Bugs in the Linux Ecosystem},
  author={Xiaochen Zou and Yu Hao and Zheng Zhang and Juefei Pu and Weiteng Chen and Zhiyun Qian},
  booktitle={NDSS},
  year={2024},
}

@INPROCEEDINGS{grebe,
  author={Lin, Zhenpeng and Chen, Yueqi and Wu, Yuhang and Mu, Dongliang and Yu, Chensheng and Xing, Xinyu and Li, Kang},
  booktitle={2022 IEEE Symposium on Security and Privacy (SP)}, 
  title={GREBE: Unveiling Exploitation Potential for Linux Kernel Bugs}, 
  year={2022},
}

@Misc{exploit1,
  Title                    = {{SSD Advisory – Linux Kernel taprio OOB}},
  HowPublished             = {\url{https://ssd-disclosure.com/ssd-advisory-linux-kernel-taprio-oob/}},
}

@Misc{exploit2,
  Title                    = {{Driving forward in Android drivers}},
  HowPublished             = {\url{https://googleprojectzero.blogspot.com/2024/06/driving-forward-in-android-drivers.html}},
}

@Misc{exploit3,
  Title                    = {{CVE-2020-27786 ( Race Condition + Use-After-Free )}},
  HowPublished             = {\url{https://ii4gsp.github.io/cve-2020-27786/}},
}

@Misc{exploit4,
  Title                    = {{A deep dive into CVE-2023-2163: How we found and fixed an eBPF Linux Kernel Vulnerability}},
HowPublished = {\url{https://bughunters.google.com/blog/6303226026131456/}%
                \url{a-deep-dive-into-cve-2023-2163-how-we-found-and-fixed-an-ebpf-linux-kernel-vulnerability}},

}

@inproceedings{llift,
  title     = {{Enhancing Static Analysis For Practical Bug Detection: An LLM-Integrated Approach}},
  author    = {Haonan Li and Yu Hao and Yizhuo Zhai and Zhiyun Qian},
  booktitle = {In Proceedings of the ACM on Programming Languages (PACMPL), Issue OOPSLA},
  year      = {2024}
}

@Misc{deferredfree,
  Title                    = {{io_uring/kbuf: defer release of mapped buffer rings}},
  HowPublished             = {\url{https://git.kernel.org/pub/scm/linux/kernel/git/torvalds/linux.git/commit/?id=c392cbecd8ec}},
}

@Misc{kernelexploitdb,
  Title                    = {{Linux Kernel Exploitation}},
  HowPublished             = {\url{https://github.com/xairy/linux-kernel-exploitation
}},
}
\bibliographystyle{abbrv}
\clearpage
\appendix

\begin{figure}[t]
    \centering
    \begin{minipage}[t]{1\linewidth}
\begin{lstlisting}[]

This patch adds to do sanity check with below fields of inode to avoid reported panic.

https:%*\textnormal{//}*)bugzilla.kernel.org/show_bug.cgi?id=200223

BUG() triggered in f2fs_truncate_inode_blocks() when un-mounting a mounted f2fs image after writing to it



// for brevity, we do not show 26 lines of messages, which are the codes of PoC

[  552.479723] F2FS-fs (loop0): Mounted with checkpoint version = 2
[  556.451891] ------------[ cut here ]------------
[  556.451899] kernel BUG at fs/f2fs/node.c:987!
// for brevity, we do not show 71 lines of messages, which are mainly the crash log
[  556.549248] BUG: KASAN: stack-out-of-bounds in arch_tlb_gather_mmu+0x52/0x170
// for brevity, we do not show 64 lines of messages, which are mainly the crash log


--- a/fs/f2fs/inode.c
+++ b/fs/f2fs/inode.c
@@ -193,9 +193,30 @@ void f2fs_inode_chksum_set(struct f2fs_sb_info *sbi, struct page *page)
 	ri->i_inode_checksum = cpu_to_le32(f2fs_inode_chksum(sbi, page));
 }
 
-static bool sanity_check_inode(struct inode *inode)
+static bool sanity_check_inode(struct inode *inode, struct page *node_page)
 {
 	struct f2fs_sb_info *sbi = F2FS_I_SB(inode);
+	unsigned long long iblocks;
+
+	iblocks = le64_to_cpu(F2FS_INODE(node_page)->i_blocks);
+	if (!iblocks) {
+		set_sbi_flag(sbi, SBI_NEED_FSCK);
+		f2fs_msg(sbi->sb, KERN_WARNING,
+			"%s: corrupted inode i_blocks i_ino=%lx iblocks=%llu, "
+			"run fsck to fix.",
+			__func__, inode->i_ino, iblocks);
+		return false;
+	}
+
+	if (ino_of_node(node_page) != nid_of_node(node_page)) {
+		set_sbi_flag(sbi, SBI_NEED_FSCK);
+		f2fs_msg(sbi->sb, KERN_WARNING,
+			"%s: corrupted inode footer i_ino=%lx, ino,nid: "
+			"[%u, %u] run fsck to fix.",
+			__func__, inode->i_ino,
+			ino_of_node(node_page), nid_of_node(node_page));
+		return false;
+	}
 
 	if (f2fs_sb_has_flexible_inline_xattr(sbi->sb)
 			\&\& !f2fs_has_extra_attr(inode)) {
@@ -267,7 +288,7 @@ static int do_read_inode(struct inode *inode)
 
 	get_inline_info(inode, ri);
 
-	if (!sanity_check_inode(inode)) {
+	if (!sanity_check_inode(inode, node_page)) {
 		f2fs_put_page(node_page, 1);
 		return -EINVAL;
 	}
\end{lstlisting}

    \end{minipage}
    \caption{A failed case by LLM component}
    \label{fig:llmfail}
\end{figure}

\subsection{Model and training details.} 
\label{sec:modeltrain}
Our experiments were on a Ubuntu 20.04.5 LTS server  with AMD EPYC 7542 processors and 4 Nvidia GeForce RTX 3080 Ti GPUs. 
The \tool model consists of 140 million parameters, including 6 encoder layers and 6 decoder layers, with a model dimension of 768 and 12 heads.
We used the Adam optimizer with a linear learning rate decay schedule for optimization, starting the training with a dropout rate of 0.1. The maximum input sequence length is set to 1024 tokens. In addition, we configured $\epsilon$ as 1e-6 and $\beta_2$ as 0.98 for the Adam optimizer.

\subsection{Manual Verification of Security Relevance.}
\label{sec:manualsecverify}

In our manual verification, we follow the definition of ``security patches" provided by GraphSPD, classifying a patch as security-related if it addresses vulnerabilities corresponding to any Common Weakness Enumeration Specification (CWE) type. Similarly, our manual process examines both the commit message and the code changes.

First, we carefully examine the commit messages for explicit or implicit indicators of security implications, such as references to ``CVE", ``use-after-free", ``double-free", ``out-of-bounds", ``memory leak", ``race condition", ``miss permission check" or terms explicitly highlighting vulnerability remediation. Commit messages containing these clear security-related keywords are immediately flagged as potential security patches.

Second, for cases lacking explicit textual indicators, we rigorously inspect the code modifications themselves, focusing especially on changes involving critical memory operations or synchronization mechanisms. Patches altering the order or presence of memory operations (e.g., allocations, frees), modifying lock handling or synchronization primitives, adjusting array indexing and bounds checking, or addressing potential uninitialized behavior and race conditions are further analyzed and typically classified as security-critical.

Conversely, patches explicitly described as introducing new features, performance improvements, or addressing purely functional issues, as well as trivial refactoring tasks or routine maintenance activities (e.g., adding simple wrappers for existing functions like \texttt{kfree()} and \texttt{kmalloc()}, or removing redundant locking), are generally categorized as non-security-related. This meticulous approach ensures robust and consistent identification of patches with genuine security implications.

\subsection{Failed cases of \system.}
\label{sec:duallmfail}

\autoref{fig:llmfail} illustrates a case where the LLM component misclassifies a patch. This patch, which fixes a memory out-of-bounds (OOB) bug, is incorrectly classified as non-UAF-OOB by the LLM.
The commit title and message together span 179 lines, including a brief introduction to the bug fix, the proof-of-concept (PoC) code, and the crash log. The introduction is minimal, mentioning only that a sanity check is added, without explicitly referencing an out-of-bounds issue.
Within the crash log, a single line—``BUG: KASAN: stack-out-of-bounds in arch\_tlb\_gather\_mmu"—clearly indicates the presence of an OOB bug. However, this critical indicator is buried within 128 lines of noisy crash information. Moreover, at the beginning of the crash log, the prominent message ``kernel BUG at fs/f2fs/node.c" may have misled the LLM into incorrectly inferring that the patch addresses a generic kernel error rather than a memory corruption vulnerability.
This case illustrates a typical LLM hallucination, where the model overlooks critical evidence within misleading or noisy context.

\autoref{fig:slicelmfail} illustrates a case where the \tool component misclassifies a patch. This patch, which fixes a memory out-of-bounds (OOB) bug, is incorrectly classified as non-UAF-OOB.
The patch spans 13,938 lines of changes, primarily involving the deletion of 17 files as part of a large-scale code cleanup. The critical code responsible for the vulnerability is deeply buried within this extensive diff and intermixed with unrelated removals.
Given the sheer volume of code changes and the lack of strong localized signals, \tool struggles to isolate the vulnerability-specific code and context needed for correct classification.
This failure aligns with the general failure mode observed in patches involving extensive modifications, in which vulnerability-related code is buried within large amounts of unrelated changes, making automated fine-grained analysis particularly challenging.

\begin{figure}[t]
    \centering
    \begin{minipage}[t]{1\linewidth}
\begin{lstlisting}[]

Lightnvm supports the OCSSD 1.x and 2.0 specs which were early attempts to produce Open Channel SSDs and never made it into the NVMe spec proper.  They have since been superceeded by NVMe enhancements such as ZNS support.  Remove the support per the deprecation schedule.


--- a/drivers/lightnvm/core.c
+++ /dev/null
// the diffs are to delete all of codes in drivers/lightnvm/core.c;for brevity, we do not show 

--- a/drivers/lightnvm/pblk-cache.c
+++ /dev/null
// the diffs are to delete all of codes in drivers/lightnvm/pblk-cache.c;for brevity, we do not show 

--- a/drivers/lightnvm/pblk-core.c
+++ /dev/null
// the diffs are to delete all of codes in drivers/lightnvm/pblk-core.c;for brevity, we do not show 

// other diffs are to delete 14 other files and other changes;for brevity, we do not show 

\end{lstlisting}

    \end{minipage}
    \caption{A failed case by \tool component}
    \label{fig:slicelmfail}
\end{figure}

\subsection{Case study for TreeVul}
\label{subsec:treevulcases}
\lstdefinelanguage{diff}{
  morecomment=[f][\color{blue}]{@@},     
  morecomment=[f][\color{red}]-,         
  morecomment=[f][\color{mygreen}]+,       
  morecomment=[f][\color{magenta}]{---}, 
  morecomment=[f][\color{magenta}]{+++},
}

\begin{figure}[t]
    \centering
    \begin{minipage}[t]{1\linewidth}
\begin{lstlisting}[]

The __mtk_foe_entry_clear() function frees "entry" so we have to use the _safe() version of hlist_for_each_entry() to prevent a use after free.

--- a/drivers/net/ethernet/mediatek/mtk_ppe.c
+++ b/drivers/net/ethernet/mediatek/mtk_ppe.c
@@ -600,6 +600,7 @@ void __mtk_ppe_check_skb(struct mtk_ppe *ppe, struct sk_buff *skb, u16 hash)
 	struct mtk_foe_entry *hwe = &ppe->foe_table[hash];
 	struct mtk_flow_entry *entry;
 	struct mtk_foe_bridge key = {};
+	struct hlist_node *n;
 	struct ethhdr *eh;
 	bool found = false;
 	u8 *tag;
@@ -609,7 +610,7 @@ void __mtk_ppe_check_skb(struct mtk_ppe *ppe, struct sk_buff *skb, u16 hash)
 	if (FIELD_GET(MTK_FOE_IB1_STATE, hwe->ib1) == MTK_FOE_STATE_BIND)
 		goto out;
 
-	hlist_for_each_entry(entry, head, list) {
+	hlist_for_each_entry_safe(entry, n, head, list) {
 		if (entry->type == MTK_FLOW_TYPE_L2_SUBFLOW) {
 			if (unlikely(FIELD_GET(MTK_FOE_IB1_STATE, hwe->ib1) ==
 				     MTK_FOE_STATE_BIND))

\end{lstlisting}
    \end{minipage}
    \caption{A patch fixing an UAF bug that TreeVul misclassifies as OOB (our method correctly classifies it)}
    \label{fig:treevul1}
\end{figure}

~\autoref{fig:treevul1} is one case that TreeVul fails but our method succeeds. TreeVul uses only commit diff as input, but this patch does not have clear features inside the commit diff to show that they fix a use-after-free bug. However, our method also considers the commit title and message, and commit titles of the patch clearly say that they fix use-after-free bugs (indicated at the top of the figure).

\begin{figure}[t]
    \centering
    \begin{minipage}[t]{1\linewidth}
\begin{lstlisting}[]

This fixes a nasty bug: dccp_send_reset() is called by both DCCPv4 and DCCPv6, but uses inet_sk_rebuild_header() in each case.This leads to unpredictable and weird behaviour: under some conditions, DCCPv6 Resets were sent, in other not.

The fix is to use the AF-independent rebuild_header routine.


--- a/net/dccp/output.c
+++ b/net/dccp/output.c
@@ -391,7 +391,7 @@ int dccp_send_reset(struct sock *sk, enum dccp_reset_codes code)
 	 * FIXME: what if rebuild_header fails?
 	 * Should we be doing a rebuild_header here?
 	 */
-	int err = inet_sk_rebuild_header(sk);
+	int err = inet_csk(sk)->icsk_af_ops->rebuild_header(sk);

 	if (err != 0)
 		return err;
   	skb = sock_wmalloc(sk, sk->sk_prot->max_header, 1, GFP_ATOMIC);
	if (skb == NULL)
		return -ENOBUFS;
	/* Reserve space for headers and prepare control bits. */
	skb_reserve(skb, sk->sk_prot->max_header);
	DCCP_SKB_CB(skb)->dccpd_type	   = DCCP_PKT_RESET;
	DCCP_SKB_CB(skb)->dccpd_reset_code = code;

	return dccp_transmit_skb(sk, skb);

\end{lstlisting}

    \end{minipage}
    \caption{A patch fixing an OOB bug that TreeVul misclassifies as NULL pointer dereference (our method correctly classifies it)}
    \label{fig:treevul3}
\end{figure}

~\autoref{fig:treevul3} is another example. The commit diff changes only one function invocation which fixes an out-of-bound (OOB) memory access bug.
The problem here is that \texttt{inet\_sk\_rebuild\_header()} builds an IPv4 header (which is smaller than an IPv6 header) regardless of the socket type. If it is expecting an IPv6 header but got an IPv4 one, the subsequent access to the packet will be out-of-bounds. The patch calls the socket's corresponding header builder function \texttt{inet\_csk(sk)->icsk\_af\_ops->rebuild\_header(sk)}, which will build either an IPv4 or an IPv6 header, depending on the type of socket. 
This patch lacks clear indications about it fixing an OOB bug. 
Specifically, TreeVul misclassifies this case as NULL pointer dereference.
Though there is no explicit hint in the commit title and message, our \tool component can capture a key statement in line 22 where 
variable \texttt{sk} is used in a memory allocation function, and a specific field of the variable \texttt{sk} determines the size of the allocated memory. Since \texttt{sk} is potentially modified due to the patch (including its field), our model correctly infers that it is fixing an OOB bug.

\subsection{List of common key phrases}
\label{sec:keyphrases}

We compiled a list of common key phrases to prepare the dataset for the step to classify memory corruption patches into memory out-of-bounds access, use-after-free and other memory-corruption patches. The commit titles are first converted to lowercase format. The key phrases for memory out-of-bounds access are ``fix out of bounds", ``fix out-of-bound", ``fix buffer overflow", ``fix stack overflow"; the key phrases for use-after-free are ``fix use after free", ``fix use-after-free"; the key phrase for other memory-corruption patches are ``fix uninit-value", ``fix uninitialize", ``fix memory leak", ``fix null pointer dereference", ``fix null dereference", ``fix null pointer reference", ``fix null-ptr deref", ``fix null-ptr-deref", ``fix null pointer access", ``fix null pointer bug", ``fix null deref", ``fix null-deref", ``fix null-ptr-deref".

\subsection{Data Cleaning of CWE Labels}
\label{sec:clean}

The bug type labels are mapped from the CWE labels. However, the direct mapping can introduce errors. First, because CWE labels follow a hierarchy, there are CVEs that received CWEs labels at the intermediate level, which does not allow us to identify its true bug type (this is also observed in prior work~\cite{pan2023fine}). For example, 5 CVEs are labeled as \texttt{CWE-20: Improper Input Validation}, but in fact, they are actually out-of-bounds bugs. Second, original CWE labels themselves can be incorrect. For example, CVE-2016-5400~\cite{wronglabelcve} is labeled as \texttt{CWE-119: Improper Restriction of Operations within the Bounds of a Memory Buffer}—a category for bugs that "reads from or writes to a memory location outside the buffer's intended boundary"~\cite{wronglabelcwe}. This label suggests an out-of-bounds bug. However, both its commit message and CVE description explicitly indicate that the bug fixed by the patch is memory leak, which is further confirmed by the patch diff showing the addition of memory free operations along an execution path. Overall, we identified and corrected 86 such cases where bug type labels directly mapped from CWE labels were incorrect.

\end{document}